\input{psfig}
\input{epsf}
\documentstyle[12pt,lathuile]{article}

\def\simge{\mathrel{%
   \rlap{\raise 0.511ex \hbox{$>$}}{\lower 0.511ex \hbox{$\sim$}}}}
\def\simle{\mathrel{
   \rlap{\raise 0.511ex \hbox{$<$}}{\lower 0.511ex \hbox{$\sim$}}}}
 
\def\slashchar#1{\setbox0=\hbox{$#1$}           
   \dimen0=\wd0                                 
   \setbox1=\hbox{/} \dimen1=\wd1               
   \ifdim\dimen0>\dimen1                        
      \rlap{\hbox to \dimen0{\hfil/\hfil}}      
      #1                                        
   \else                                        
      \rlap{\hbox to \dimen1{\hfil$#1$\hfil}}   
      /                                         
   \fi}                                         %
\def\nn{\nonumber}
\def\ts{\thinspace}

\def\ra{\rightarrow}

\def\ol{\bar}

\def\be{\begin{equation}} 
\def\ee{\end{equation}} 
\def\bea{\begin{eqnarray}}
\def\eea{\end{eqnarray}}
\def\ba{\begin{array}}
\def\ea{\end{array}}
\def\chipr{\chi^{\ts \prime}}

\def\CO{{\cal O}}

\def\ecm{\sqrt{s}}

\def\atc{\alpha_{TC}}

\def\atro{\alpha_{\tro}}

\def\Ntc{N_{TC}}
\def\suc{SU(3)}

\def\sutc{SU(\Ntc)}

\def\getc{g_{ETC}}

\def\thw{\theta_W}
\def\kslash{\raise.15ex\hbox{/}\kern-.57em k}
\def\LTC{\Lambda_{TC}}

\def\METC{M_{ETC}}
\def\tro{\rho_{T}}
\def\tros{\rho_{T8}^{0}} 
\def\troct{\rho_{T8}} 
\def\tropm{\rho_{T}^\pm}

\def\troz{\rho_{T}^0}
\def\tom{\omega_T}
\def\tpi{\pi_T}
\def\tpipm{\pi_T^\pm}
\def\tpimp{\pi_T^\mp}
\def\tpip{\pi_T^+}
\def\tpim{\pi_T^-}
\def\tpiz{\pi_T^0}
\def\tpipr{\pi_T^{0 \prime}}

\def\octpi{\pi_{T8}}
\def\octpipm{\pi_{T8}^\pm}

\def\octpiz{\pi_{T8}^0}

\def\condtc{{\langle \ol T T \rangle}_{TC}}
\def\condetc{{\langle \ol T T \rangle}_{ETC}}
\def\tpilq{\pi_{\ol L Q}}

\def\tpind{\pi_{\ol N D}}
\def\tpied{\pi_{\ol E D}}
\def\tpiql{\pi_{\ol Q L}}

\def\jet{{\rm jet}}
\def\jets{{\rm jets}}

\def\mev{{\rm MeV}}
\def\gev{{\rm GeV}}
\def\tev{{\rm TeV}}

\def\ipb{{\rm pb}^{-1}}

\def\ifb{{\rm fb}^{-1}}
\def\half{{\textstyle{ { 1\over { 2 } }}}}

\begin{document}
\title{ 
TECHNICOLOR SIGNATURES---IERI, OGGI E DOMANI
}
\author{
KENNETH LANE\\
{\em Department of Physics, Boston University}\\
{\em 590 Commonwealth Avenue, Boston, Massachusetts 02215} \\
}
\maketitle
\baselineskip=14.5pt
\begin{abstract}
We briefly review the basic structure of modern--day technicolor theories. We
then discuss the signatures for technicolor as they were in its early days,
the searches that have been recently performed at the Tevatron and LEP
colliders, and the prospects for testing technicolor in Tevatron Run~II and
at the LHC.
\end{abstract}
\baselineskip=17pt
\newpage

\section*{1. Introduction}

Since the standard model was constructed in the first half of the 1970s,
there has been discomfort and dissatisfaction with one of its foundations,
the description of electroweak and flavor symmetry breaking in terms of one
or more elementary Higgs boson multiplets. This is no description at all.  No
dynamical reason is provided for electroweak symmetry breaking. There is no
understanding of why its energy scale is roughly 1~TeV and, in particular why
it is so much less than the GUT scale, if there is one, or the Planck scale
if there isn't. This is the hierarchy problem. Another difficulty is the
naturalness problem.\cite{natural} Why should the Higgs mass, which suffers
quadratic renormalization, be very much less than the natural cutoff of the
theory, the GUT scale or the Planck scale?  Furthermore, in all elementary
Higgs models, every aspect of flavor---from the primordial symmetry that
tells us the number of quark and lepton generations to the weird pattern of
flavor breaking---is completely arbitrary, put in by hand. Finally, it is now
well understood that elementary Higgs models are ``trivial'', i.e., they
become free field theories in the limit of infinite cutoff.\cite{trivial}
This means that these models are at best effective, describing a more
fundamental theory that must be used above some finite energy. If the Higgs
boson is light, less than 200--300~GeV, this transition to a more fundamental
theory may be postponed until very high energy, but what lies up there
worries us nonetheless.

In response to these shortcomings, the dynamical picture of electroweak and
flavor symmetry breaking emerged in 1978--80. This picture, now known as
technicolor (TC)~\cite{tc,kltasi,rscreview,frascati} and extended technicolor
(ETC),\cite{etceekl,etcsd} was motivated first of all by the premise that
{\it every} fundamental energy scale should have a dynamical origin. Thus,
the weak scale (or vacuum expectation value of the Higgs field) $v = 2^{-1/4}
G_F^{-1/2} = 246\,\gev$ should reflect the fundamental scale of a new strong
dynamics, technicolor, just as the pion decay constant $f_\pi = 93\,\mev$
reflects QCD's scale of about 200~MeV. For this reason, I write $F_\pi =
2^{-1/4} G_F^{-1/2} = 246\,\gev$ to emphasize its dynamical origin.

Technicolor, a gauge theory of fermions with no elementary scalars, is
modeled on the precedent of QCD. In QCD, massless quarks have a chiral
symmetry that is spontaneously broken, giving rise to massless Goldstone
bosons, the pions. When this happens in technicolor, three Goldstone bosons
become, via the Higgs mechanism, the longitudinal components $W_L^\pm$ and
$Z_L^0$ of the weak bosons. With technifermions forming left--handed doublets
and right--handed singlets under electroweak $SU(2)\otimes U(1)$, the
electroweak masses are $M_W = M_Z\cos\thw = \half g F_\pi$, where $g =
e/\sin\thw$.

Like QCD, technicolor is asymptotically free. This solves in one stroke the
naturalness, hierarchy, and triviality problems. If we imagine that the
technicolor gauge symmetry (taken here to be $\sutc$) is embedded at a very
high energy $\Lambda$ in some grand unified gauge group, then TC's
characteristic scale $\LTC$, where the coupling $\atc$ becomes strong enough
to trigger chiral symmetry breaking, is naturally exponentially smaller than
$\Lambda$. The mass of all technihadrons, including Higgs--like scalars
(though that language is neither accurate nor useful in technicolor) is of
order $\LTC$. And asymptotically free field theories are nontrivial. No other
scenario for the physics of the TeV scale solves these problems so
neatly. Period.

Technicolor alone cannot address the flavor problem. Something more is needed
to communicate electroweak symmetry breaking to quarks and leptons and give
them mass. Furthermore, in all but the minimal TC model with a single doublet
of technifermions, there are Goldstone bosons---technipions $\tpi$, in
addition to $W_L^\pm$ and $Z_L^0$---that must be given mass. Their masses must
be more than 50--100~GeV for these $\tpi$ to have escaped detection.
Extended technicolor was invented to address these and other aspects of
flavor physics. It was also motivated by the desire to make flavor
understandable at energies well below the GUT scale in terms of gauge
dynamics of the kind that worked so neatly for electroweak symmetry breaking,
namely, technicolor.

In extended technicolor, the ordinary $SU(3)$ color, $\sutc$ technicolor, and
flavor symmetries are unified into a larger ETC gauge group. This symmetry is
broken down at a scale of 100s of TeV into $SU(3) \otimes \sutc$. The ETC
interactions must break explicitly all {\it global} flavor symmetries. The
broken gauge interactions are mediated by massive ETC boson exchange and they
connect technifermions to each other, giving mass to technipions, and to
quarks and leptons, giving mass to them. Generic expressions for technipion
masses in terms of four--technifermion condensate renormalized at the ETC
scale are
\be\label{eq:tpimass}
F^2_T M^2_{\pi_T}  \simeq 2\ts {g^2_{ETC} \over {M^2_{ETC}}}
\ts \langle \ol T_L T_R \ol T_R T_L \rangle_{ETC} \ts.
\ee
Here, $F_T$ is the technipion decay constant. In TC models containing $N$
doublets of color--singlet technifermions, $F_T = F_\pi/\sqrt{N}$. Typical
quark and lepton masses are given by
\be\label{eq:qlmass}
m_q(\METC) \simeq m_\ell(\METC)  \simeq {g_{ETC}^2 \over
{2 M_{ETC}^2}} \condetc \ts,
\ee
where $\condetc$ is the bilinear technifermion condensate renormalized at
$\METC$. This is related to the condensate renormalized at
$\LTC$, expected by scaling from QCD to be
\be\label{eq:ctc}
\condtc \simeq 4 \pi F^3_T \ts,
\ee
by
\be\label{eq:condrenorm}
\condetc = \condtc \ts \exp\left(\int_{\LTC}^{M_{ETC}} \ts {d \mu
\over {\mu}} \ts \gamma_m(\mu) \right) \ts.
\ee
The anomalous dimension $\gamma_m$ of the operator $\ol T T$ is given in
perturbation theory by
\be\label{eq:gmm}
\gamma_m(\mu) = {3 C_2(R) \over {2 \pi}} \atc(\mu) + O(\atc^2) \ts,
\ee
where $C_2(R)$ is the quadratic Casimir of the technifermion
$\sutc$--representation $R$. For the fundamental representation of $\sutc$,
$C_2(\Ntc) = (\Ntc^2-1)/2\Ntc$.

Extended technicolor, like {\it all} other attempts to understand flavor, has
met many obstacles and there is still no ``standard ETC'' model. This is
perhaps not surprising since ETC is essentially a problem in strongly coupled
dynamics. Thus, from the beginning, we (few) technicolor enthusiasts have
pushed hard for experimental tests of its basic ideas. In the rest of this
paper, I will review and preview those searches for technicolor.

\section*{2. Yesterday}
\begin{figure}[t]
\vspace{9.0cm}
\includegraphics{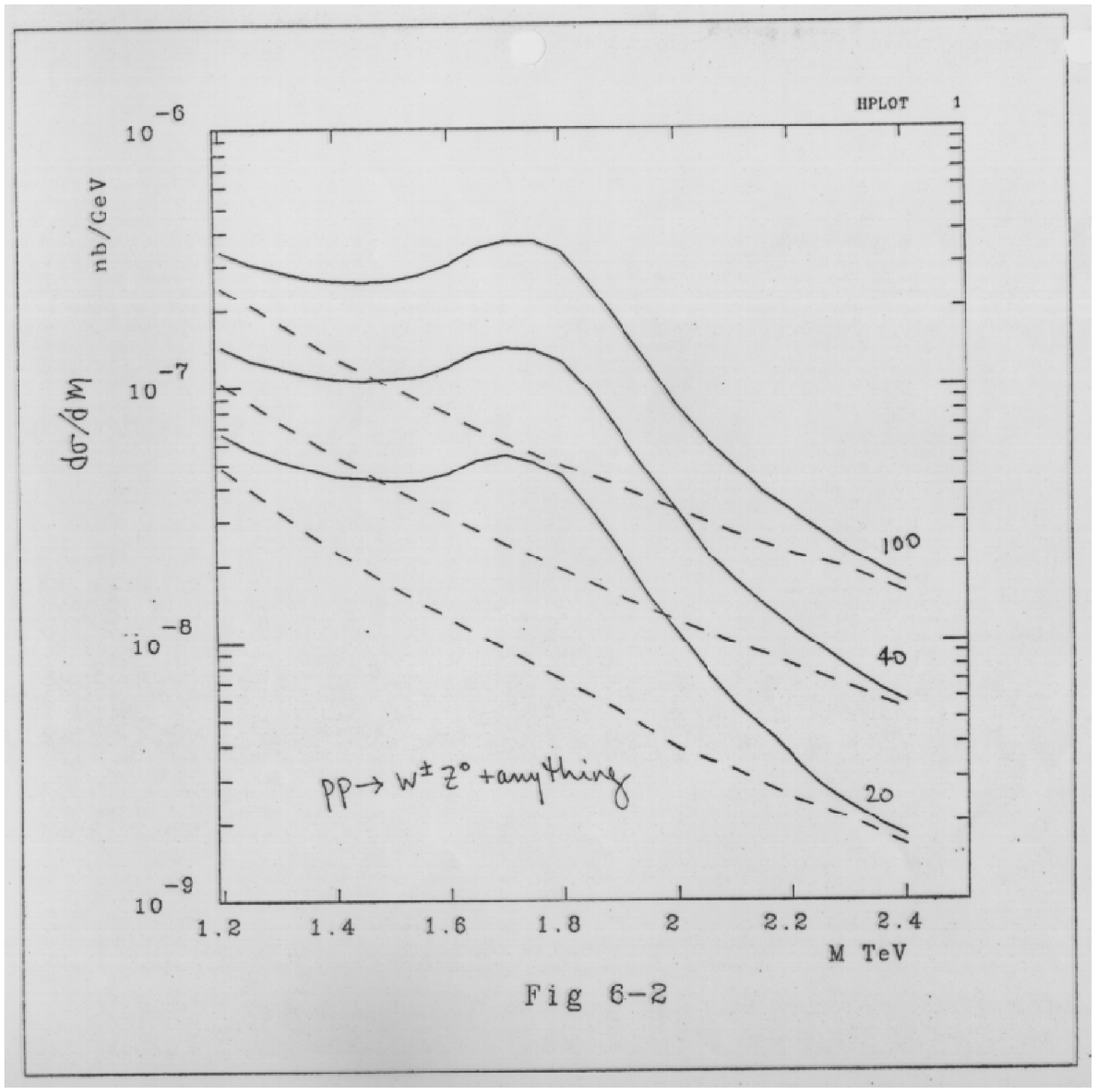}
\vskip3.0truecm
 \caption{\it
      Mass spectrum in $pp$ collisions at $\ecm = 20$, $40$, $100\,\tev$ for
      $\rho^\pm_T \ra W^\pm Z^0$ with $M_{\tro} = 1.75\,\tev$. Dashed lines
      show the standard model contribution; from Ref.~[9].
    \label{fig1} }
\end{figure}
\begin{figure}[t]
 \vspace{9.0cm}
\includegraphics{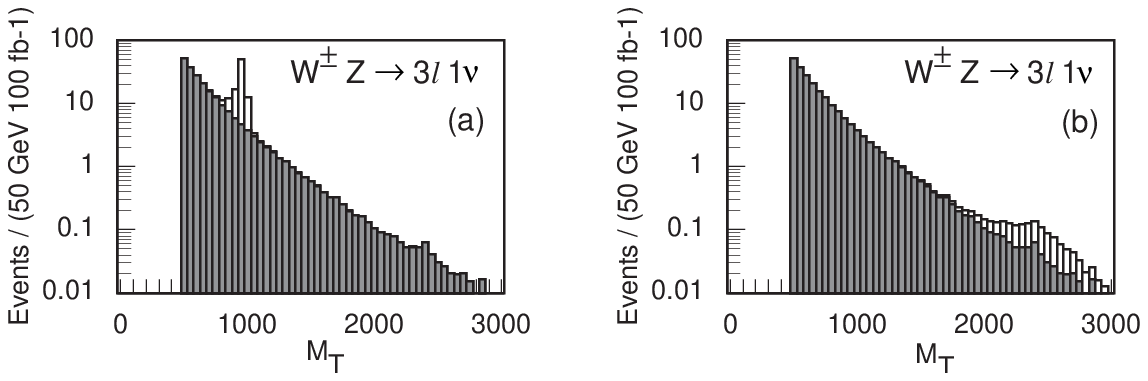}
\vskip-4.0truecm
 \caption{\it
      Event yields at the LHC for $\tropm \ra W^\pm Z^0 \ra \ell^\pm
      \nu_\ell \ell^+ \ell^-$ for $M_{\tro} = 1.0$, $2.5\,\tev$; from
      Ref.~[10].
    \label{fig2} }
\end{figure}
\begin{figure}[t]
 \vspace{9.0cm}
\includegraphics{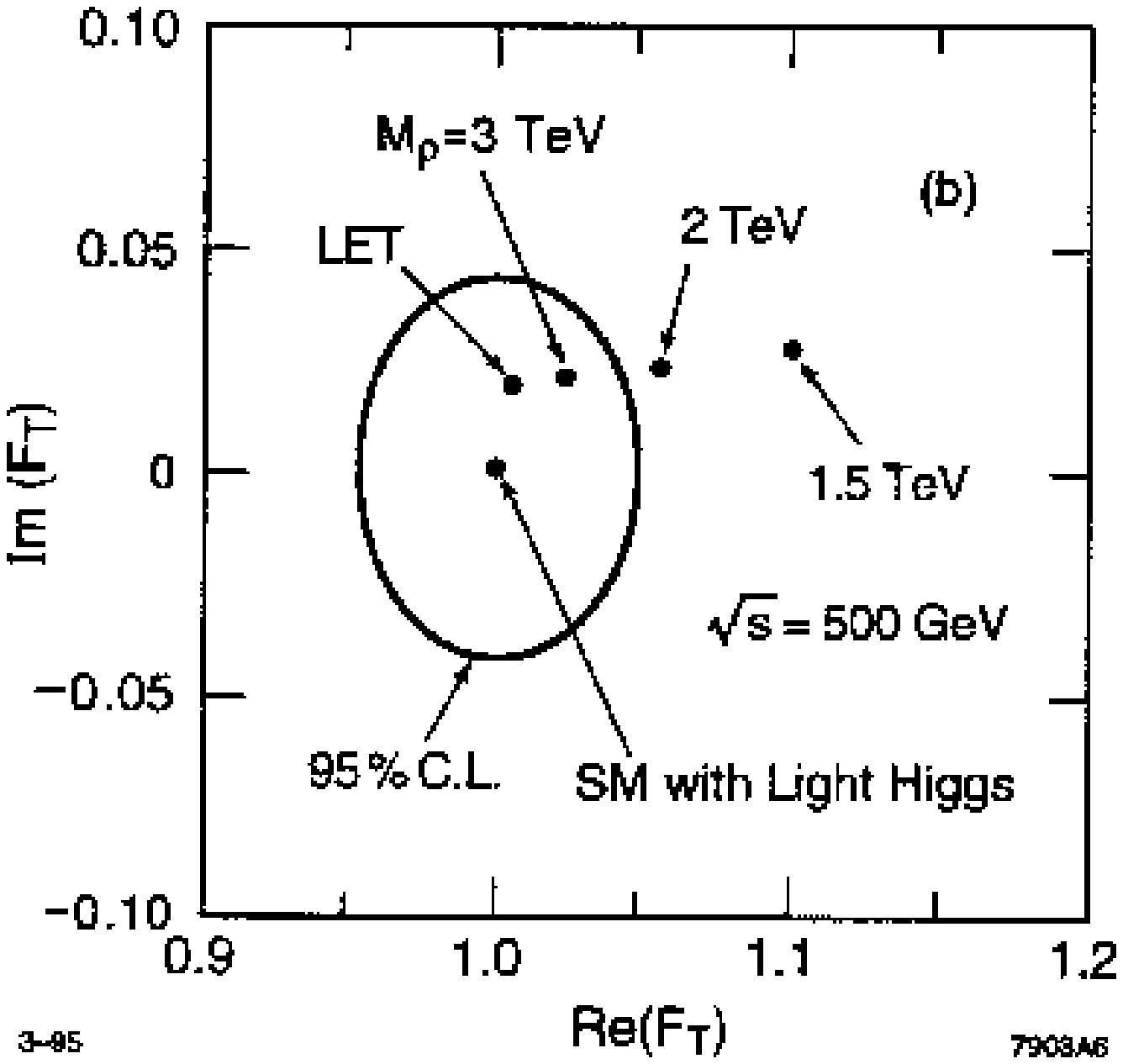}
\vskip2.2truecm
 \caption{\it
      Sensitivity of a 500~GeV NLC to $M_{\tro}$ via the $W$--boson form
      factor; from Ref.~[11].
    \label{fig3} }
\end{figure}

At the very beginning of technicolor, Susskind proposed one of its most
enduring signals.\cite{tc} In any model of technicolor, one expects bound
technihadrons with a spectrum of mesons paralleling what we see in QCD. These
will include spin--zero technipions and spin--one isovector technirhos and
isoscalar techniomegas. In the minimal one--technidoublet model ($T =
(T_U,T_D)$), the technipions are, by virtue of the Higgs mechanism, the
longitudinal compononents $W_L$ of the massive weak gauge bosons. Susskind
pointed out that the analog of the QCD decay $\rho \ra \pi\pi$ is $\tro \ra
W_L W_L$. In the limit that $M_{\tro} \gg M_{W,Z}$, the equivalence theorem
states that the amplitude for $\tro \ra W_L W_L$ has the same form as the one
for $\rho \ra \pi\pi$. If one scales TC from QCD using large--$\Ntc$
arguments, it is easy to estimate the strength of this amplitude and the
$\tro$ mass and decay rate~\cite{ehlq}:
\bea\label{eq:minrhot}
&&M_{\tro} = \sqrt{{3\over{\Ntc}}}\ts {F_\pi\over{f_\pi}} M_\rho
\simeq 2 \sqrt{{3\over{\Ntc}}}\,\tev \ts, \nn\\
&&\Gamma(\tro \ra W_L W_L ) = {2\atro p_W^3\over{3 M^2_{\tro}}} \simeq 500
\left({3\over{N_{TC}}} \right)^{3/2}\,\gev  \ts.
\eea
Here, the naive scaling argument gives $\atro = (3/\Ntc) \alpha_\rho$ where
$\alpha_\rho = 2.91$.

For a long time this was the ``benchmark'' signature for technicolor---the
analog of the search for the standard model Higgs for $M_H \simeq
100$--$800\,\gev$---and some effort has gone into establishing the reach of
hadron and lepton colliders for this process. The first accurate calculation
(i.e., with standard model interference) was carried out at the parton level
in EHLQ for the Superconducting Super Collider with $\ecm =
10$--$100\,\tev$.\cite{ehlq} Unfortunately, the SSC was cancelled before
detailed (particle level) simulations could be carried out and all that was
done is exemplified in Fig.~1 for $\tro^\pm \ra W^\pm Z^0$ with $M_{\tro} =
1.75\,\tev$ (i.e., $\Ntc = 4$). CERN's Large Hadron Collider has no reach for
such a heavy $\tro$, as can be seen in Fig.~2.\cite{golden} That, in fact, is
why the 40~TeV energy and $10^{33}$--$10^{34}$ luminosity were chosen for the
SSC.

The high energy lepton collider that has studied its reach for $\troz \ra
W^+_L W^-_L$ is the Next Linear Collider.\cite{barklow} As currently
designed, the NLC reaches to 500~GeV, extendible to 1~TeV. Having no direct
reach for a minimal technicolor $\tro$ with the expected mass, the NLC relies
on probing the form factor in $W^+W^-$ production at $s \ll M^2_{\tro}$. In
Fig.~3, for a linear collider with $\ecm = 500\,\gev$, a reach up to
$M_{\tro} \simeq 2\,\tev$ at the 95\% confidence level is indicated. This is
about as heavy as one would expect a minimal--model ground state technirho to
be. Of course, one would not be satisfied until it were directly observed as
an $s$--channel resonance. Since we still do not know the dynamics underlying
electroweak symmetry breaking, this is one of the reasons that the NLC (which
will come on well after the LHC anyway) needs to have an energy of at least
1.5--2~TeV. In any case, the 40~TeV high--luminosity SSC was and still is the
right collider to build. Alas, that is politically impossible.

It is possible that, like the search for the minimal standard model Higgs
boson, all this emphasis on the $W_L W_L$ decay mode of the $\tro$ is
somewhat misguided.\cite{kltasi} Since the minimal $\tro$ is so much heavier
than $2M_W$, this decay mode may be suppressed by the high $W$--momentum in
the decay form factor. Then, $\tro$ decays to four or more weak bosons may be
competitive or even dominate. This means that the minimal $\tro$ may be wider
than indicated in Eq.~(\ref{eq:minrhot}) and, in any case, that its decays
are much more complicated than previously thought. Furthermore, walking
technicolor,\cite{wtc} discussed below, implies that the spectrum of
technihadrons cannot be exactly QCD--like. Rather, there must be something
like a tower of technirhos extending almost up to $\METC \simge$ several
100~TeV. Whether or not these would appear as discernible resonances is an
open question.\cite{hemc} All these remarks apply as well to the isoscalar
$\tom$ and its excitations.

As everyone knows, technicolor and extended technicolor are challenged by
flavor--changing neutral current interactions (FCNC),\cite{etceekl} by
precision measurements of electroweak quantities (STU),\cite{pettests} and by
the large mass of the top quark. In general, we expect four--quark contact
interactions generated by ETC exchange. Even if these are
generation--conserving to start with, quark mixing is bound to result in
$\vert \Delta S \vert = 2$ terms with strength $g^2_{ETC} V^2_{ds}/\METC^2$.
Here, $V_{ds}$ is a mixing angle factor, presumed to be of order 0.1. The
$K_L$--$K_S$ mass difference and the CP--violating parameter $\epsilon$ imply
the constraints~\cite{etceekl,kltasi,frascati}
\be\label{eq:fcnc}
{M_{ETC} \over {g_{ETC} \ts \sqrt{{\rm Re}(V^2_{ds})}}} \simge
1300\,\tev \ts,\qquad
{M_{ETC} \over {g_{ETC} \ts \sqrt{{\rm Im}(V^2_{ds})}}} \simge
16000\,\tev \ts.
\ee
If we naively scale the technifermion condensates in
Eqs.~(\ref{eq:tpimass},\ref{eq:qlmass})from QCD, i.e., assume the anomalous
dimension $\gamma_m$ is small so that $4 \langle \ol T_L T_R \ol T_R T_L
\rangle_{ETC} \simeq \condetc^2 \simeq \condtc^2 \simeq (4 \pi F^3_T)^2$, we
obtain technipion and quark and lepton masses that are at least 10--1000
times too small, depending on the size of $V_{ds}$. This is the FCNC
problem. It is remedied by the non--QCD--like dynamics of technicolor with a
slowly running gauge coupling, ``walking technicolor'', which will be
described in the next section.

Precision electroweak measurements actually challenge technicolor, not
extended technicolor. The most cited constraint involves the so--called $S$
parameter whose measured value is $S = -0.07 \pm 0.11$ (for $M_H =
100\,\gev$).\cite{pdg} the value obtained in technicolor by scaling from QCD
is $\CO(1)$.  For example, for $N$ color--singlet technidoublets, Peskin and
Takeuchi found~\cite{pettests}
\be\label{eq:svalue}
S = 4 \pi \left(1 + {M^2_{\rho_T} \over{M^2_{a_{1T}}}}\right ) {F^2_\pi \over
{M^2_{\rho_T}}} \simeq 0.25 \ts N {N_{TC}\over{3}} \ts.
\ee
The resolution to this problem may also be found in walking technicolor. One
thing is sure: naive scaling of $S$ from QCD is unjustified and probably
incorrect in walking gauge theories. No reliable estimate exists because no
data on walking gauge theories is available to put into $S$'s calculation.

The large top quark mass requires a different dynamical innovation than
walking technicolor. Extended technicolor cannot explain the top quark's mass
without running afoul of either experimental constraints from the parameter
$\rho = M^2_W/M^2_Z\cos^2\thw$ and the $Z \ra \ol b b$ decay
rate\cite{zbbth}---the ETC mass must be about 1~TeV to produce $m_t =
175\,\gev$; see Eq.~(\ref{eq:qlmass})---or of cherished notions of
naturalness---$\METC$ may be higher, but the coupling $\getc$ then must be
fine-tuned near to a critical value. The best idea to explain the top mass so
far is topcolor--assisted technicolor,\cite{tctwohill} in which a new gauge
interaction, topcolor,\cite{topcref} becomes strong near 1~TeV and generates
a large $\ol t t$ condensate and top mass. This, too, will be described in
the next section.

\begin{figure}[t]
 \vspace{9.0cm}
\includegraphics{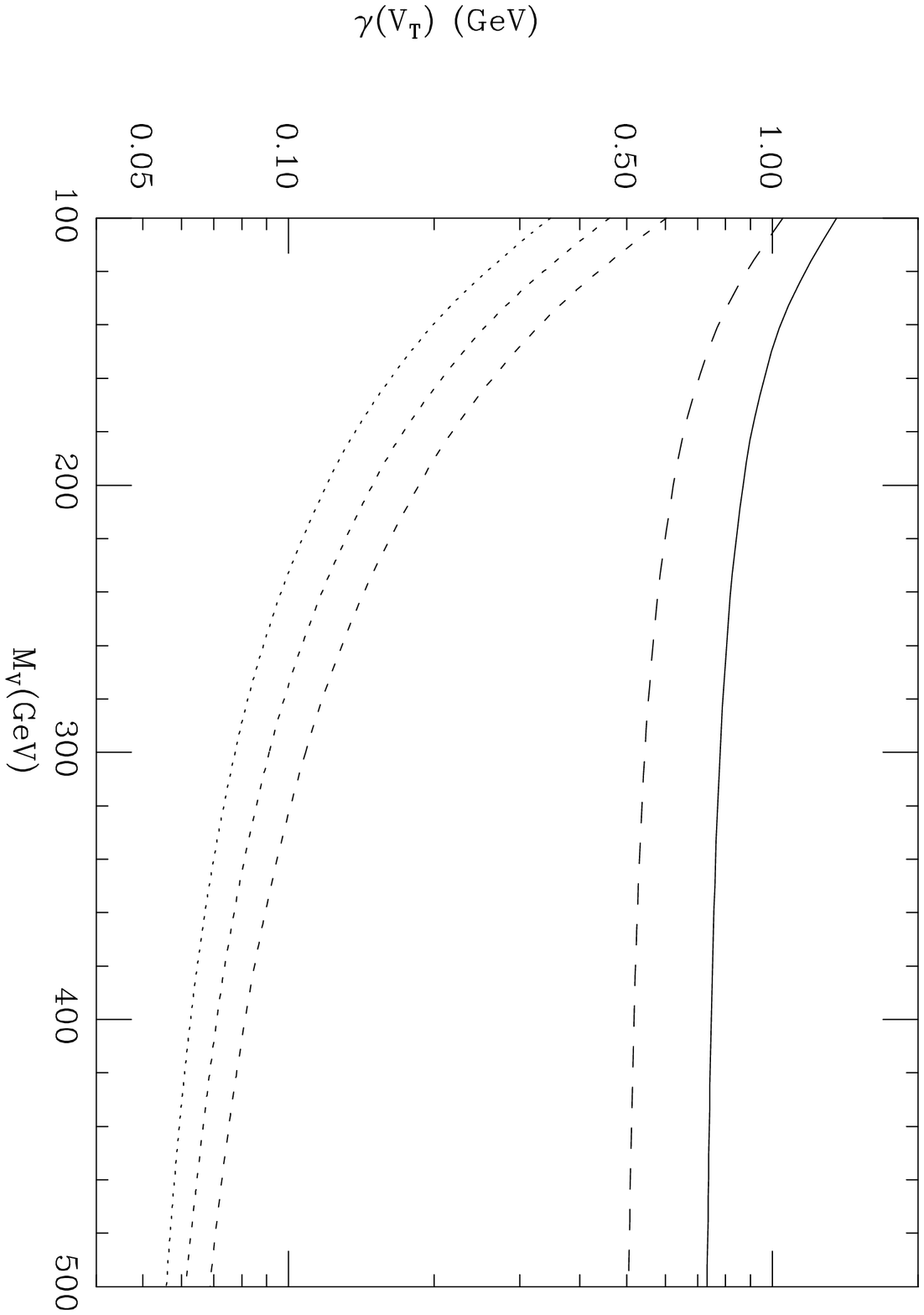}
\vskip2.5truecm
 \caption{\it
      Technivector meson decay rates versus $M_V = M_A$ for $\troz$ (solid
  curve) and $\tropm$ (long-dashed) with $M_{\tro} = 210\,\gev$, and $\tom$
  with $M_{\tom} = 200$ (lower dotted), 210 (lower short-dashed), and
  $220\,\gev$ (lower medium-dashed); $Q_U + Q_D = 5/3$ and $M_{\tpi} =
  110\,\gev$; from Ref.~[26].
    \label{fig4} }
\end{figure}
\begin{figure}[t]
 \vspace{9.0cm}
\includegraphics{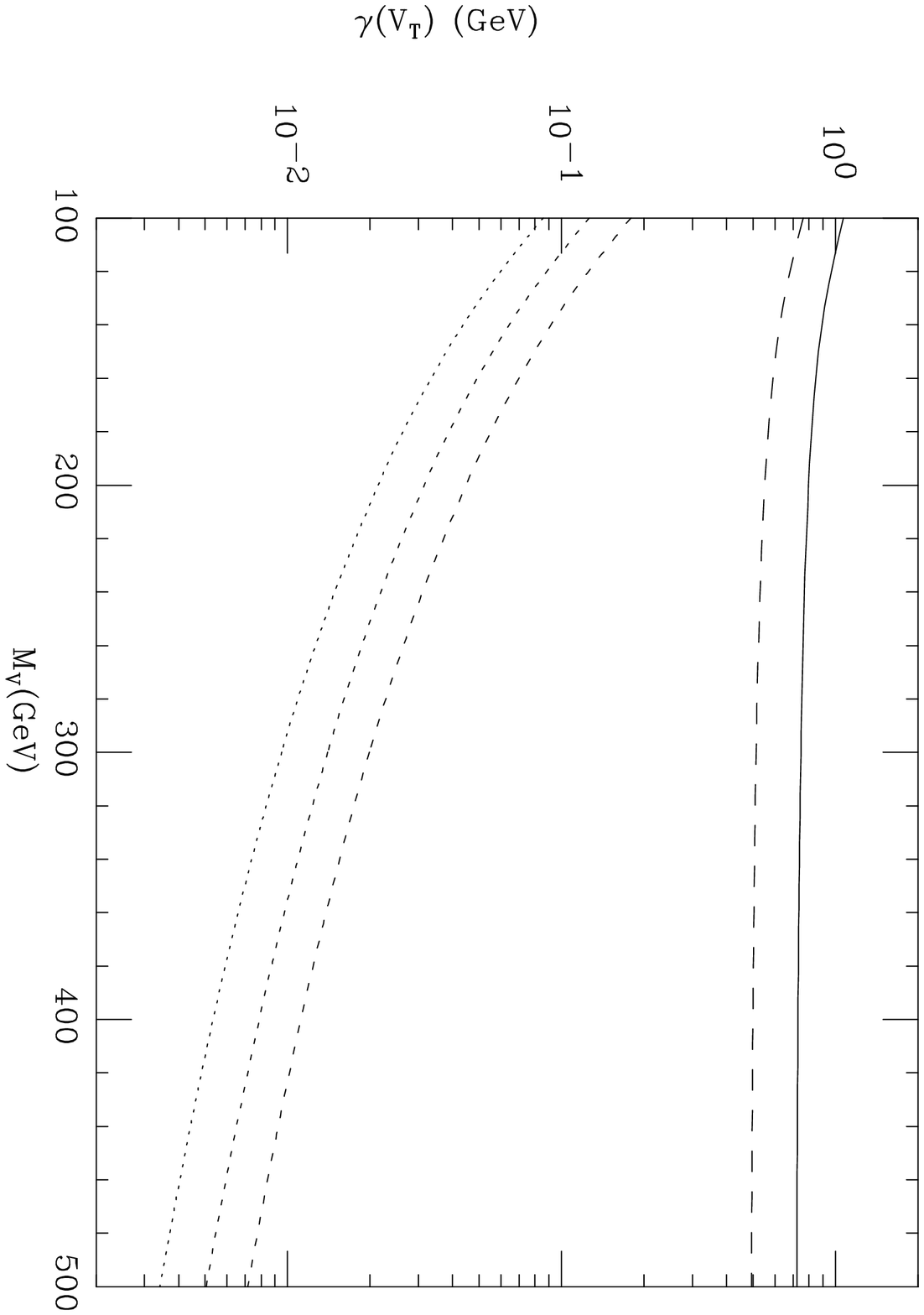}
\vskip2.5truecm
 \caption{\it
      Decay rates as in Fig.~4, with $Q_U + Q_D = 0$; from Ref.~[26].
    \label{fig5} }
\end{figure}

\begin{figure}[t]
 \vspace{9.0cm}
\includegraphics{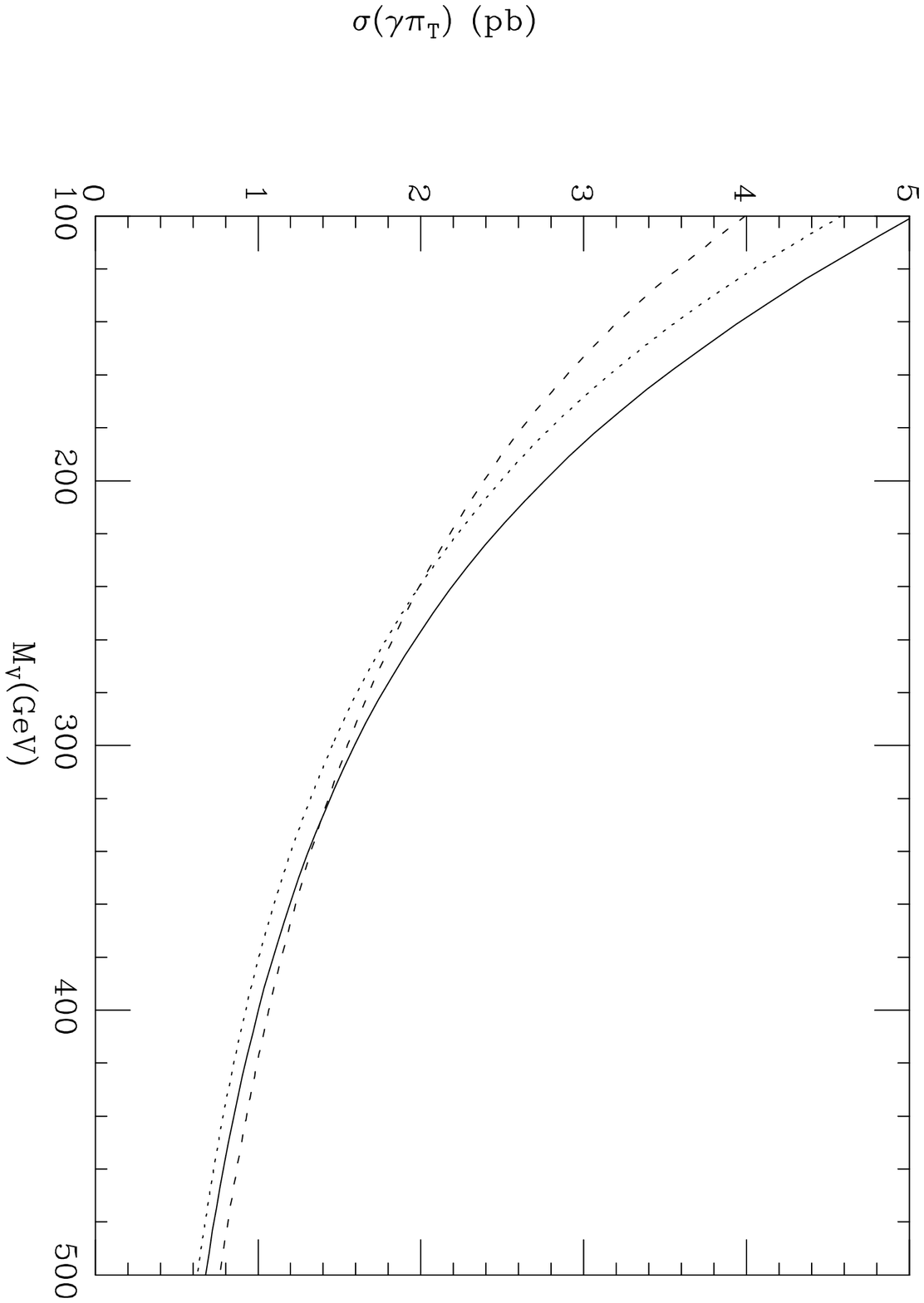}
\vskip2.5truecm
 \caption{\it
   Production rates in $p \ol p$ collisions at $\ecm = 2\,\tev$
   for the sum of $\tom$, $\troz$, $\tropm \ra \gamma
   \tpi$ versus $M_V$, for $M_{\tro} = 210\,\gev$ and $M_{\tom} = 200$ (dotted
   curve), 210 (solid), and $220\,\gev$ (short-dashed); $Q_U + Q_D = 5/3$, and
   $M_{\tpi} = 110\,\gev$; from Ref.~[26].
   \label{fig6} }
\end{figure}
\begin{figure}[t]
 \vspace{9.0cm}
\includegraphics{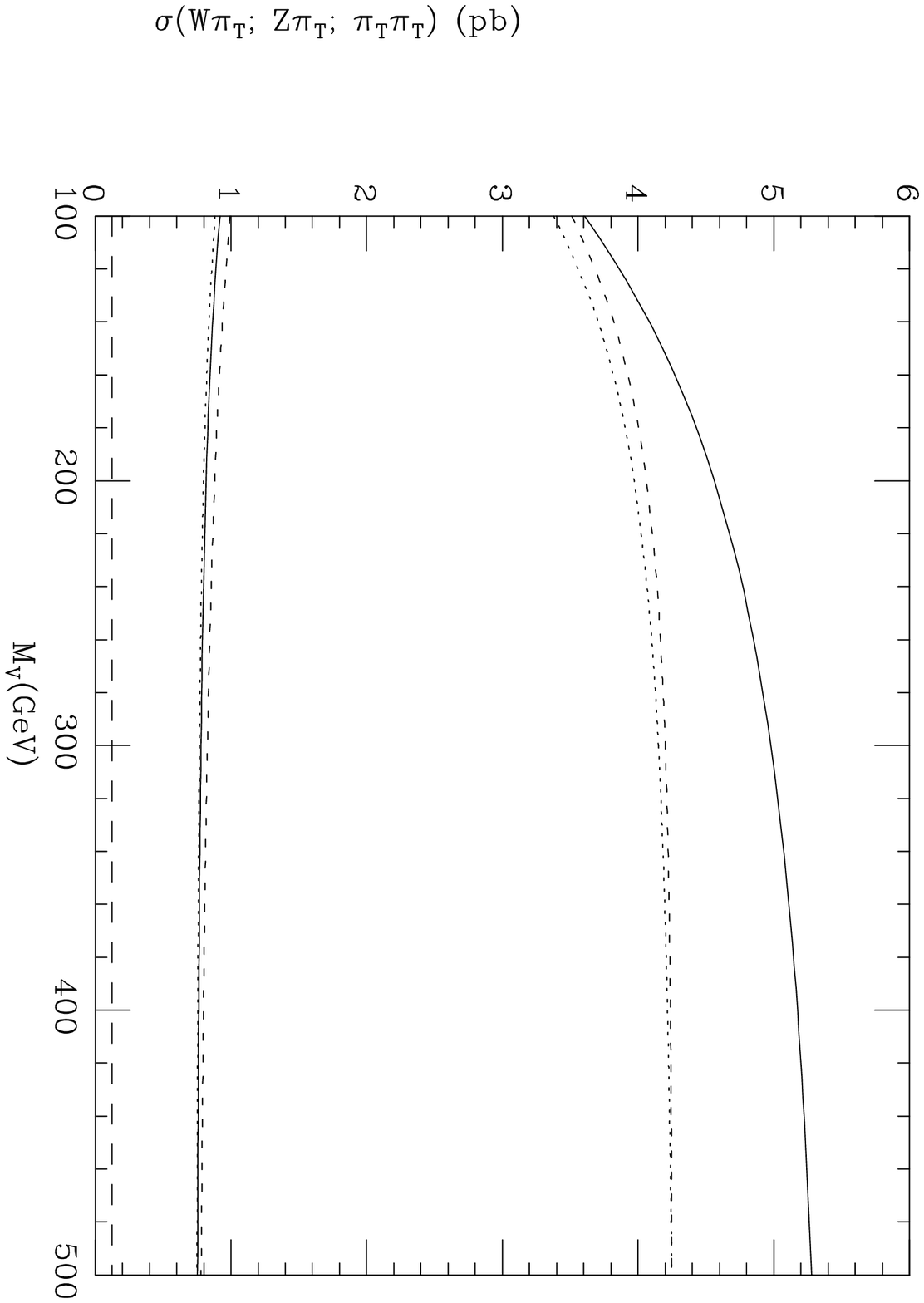}
\vskip2.5truecm
 \caption{\it
   Production rates for $\tom$, $\troz$, $\tropm \ra W \tpi$ (upper
   curves) and $Z\tpi$ (lower curves) versus $M_V$, for $M_{\tro} = 210\,\gev$
   and $M_{\tom} = 200$ (dotted curve), 210 (solid), and $220\,\gev$
   (short-dashed); $Q_U + Q_D = 5/3$ and $M_{\tpi} = 110\,\gev$. Also shown
   is $\sigma(\tro \ra \tpi\tpi)$ (lowest dashed curve); from Ref.~[26].
    \label{fig7} }
\end{figure}

\begin{figure}[t]
 \vspace{9.0cm}
\includegraphics{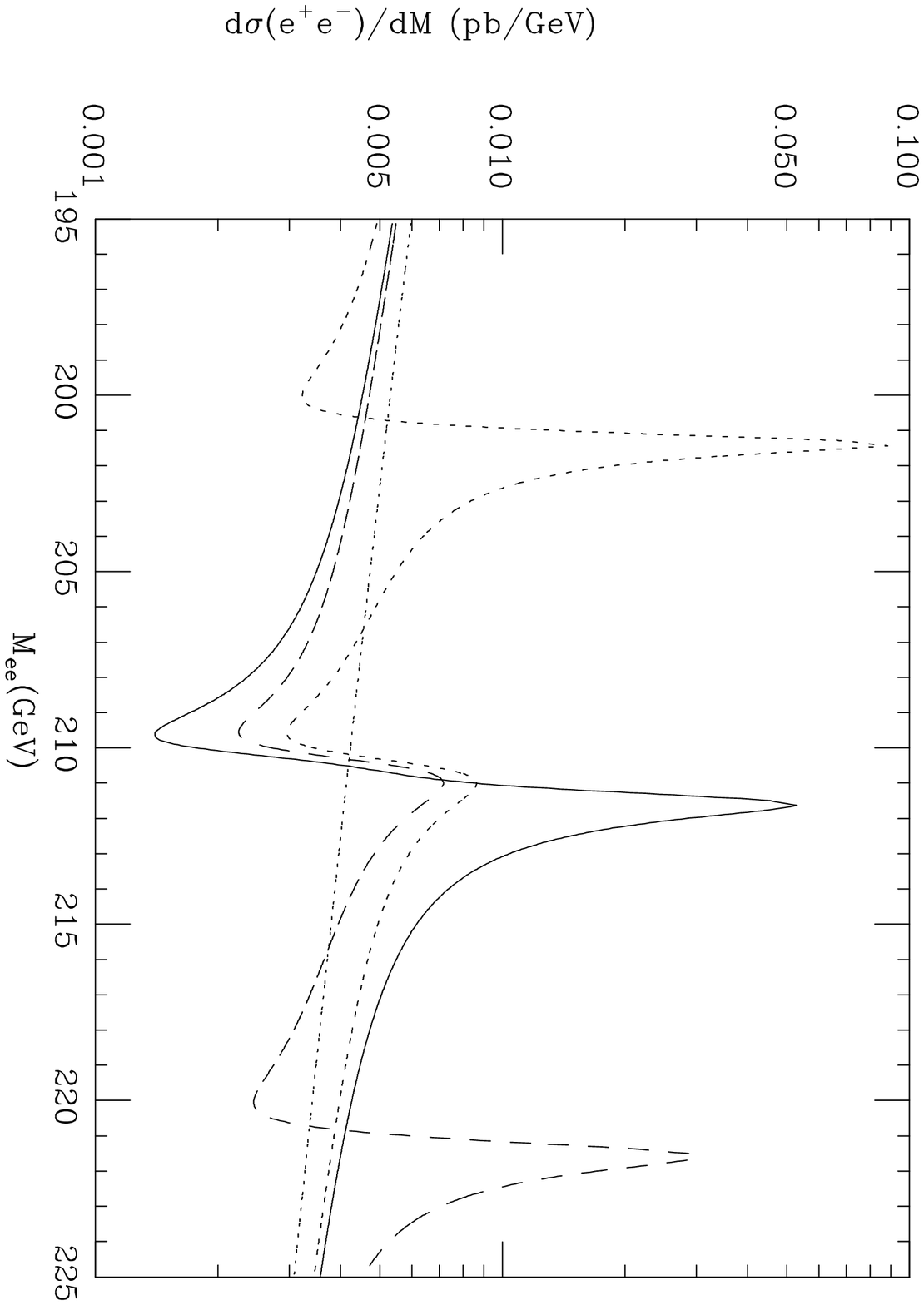}
\vskip2.5truecm
 \caption{\it
   Invariant mass distributions for $\tom$, $\troz \ra e^+e^-$
   for $M_{\tro} = 210\,\gev$ and $M_{\tom} = 200$ (short-dashed curve), 210
   (solid), and $220\,\gev$ (long-dashed); $M_V = 100\,\gev$. The standard
   model background is the sloping dotted line. $Q_U + Q_D = 5/3$ and
   $M_{\tpi} = 110\,\gev$; from Ref.~[26].
    \label{fig8} }
\end{figure}
\begin{figure}[t]
 \vspace{9.0cm}
\includegraphics{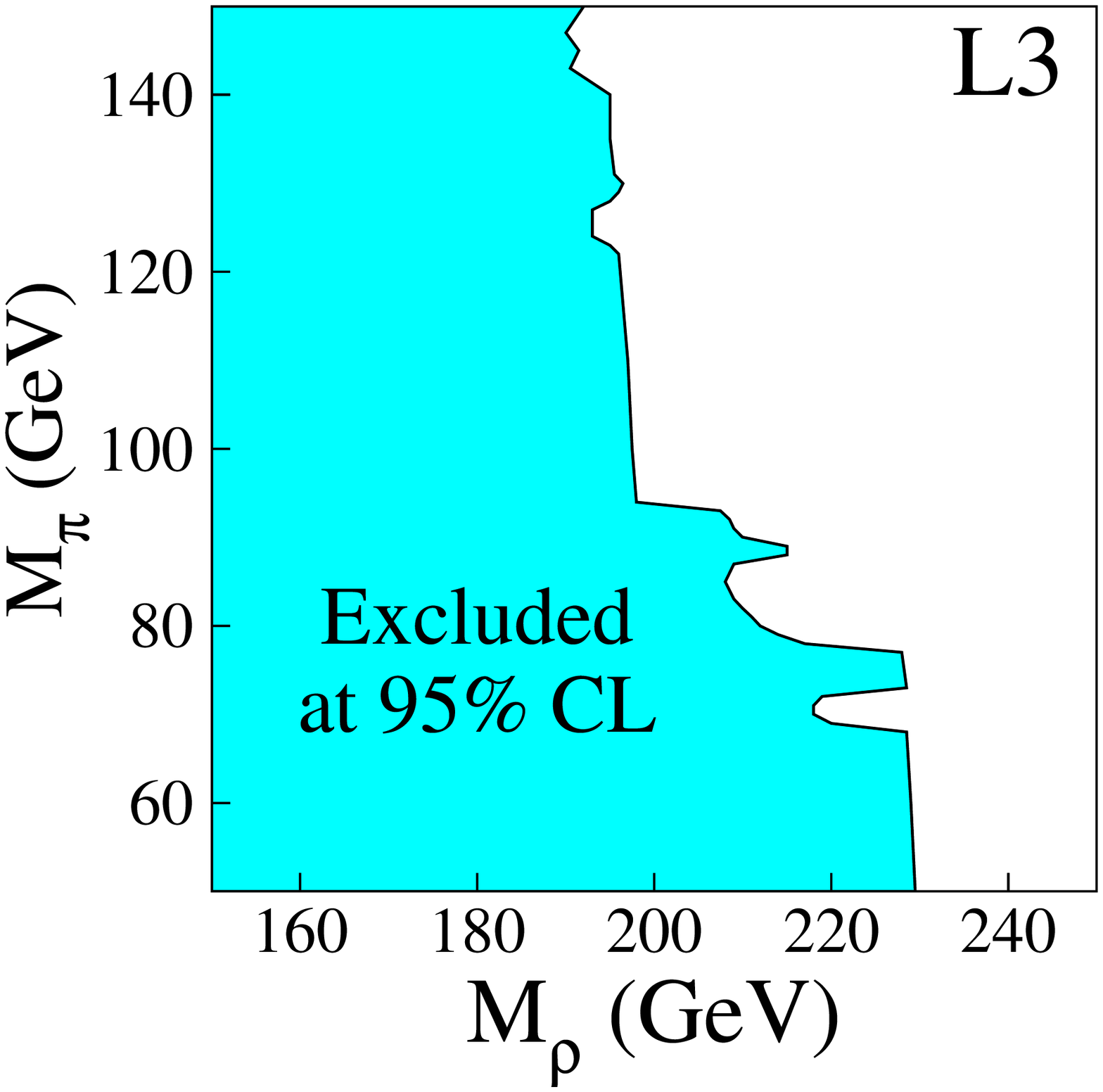}
\vskip3.0truecm
 \caption{\it
      The $M_{\tro}$--$M_{\tpi}$ region excluded by L3 at the 95\% CL; from
      Ref.~[30].
    \label{fig9} }
\end{figure}

\begin{figure}[t]
 \vspace{9.0cm}
\includegraphics{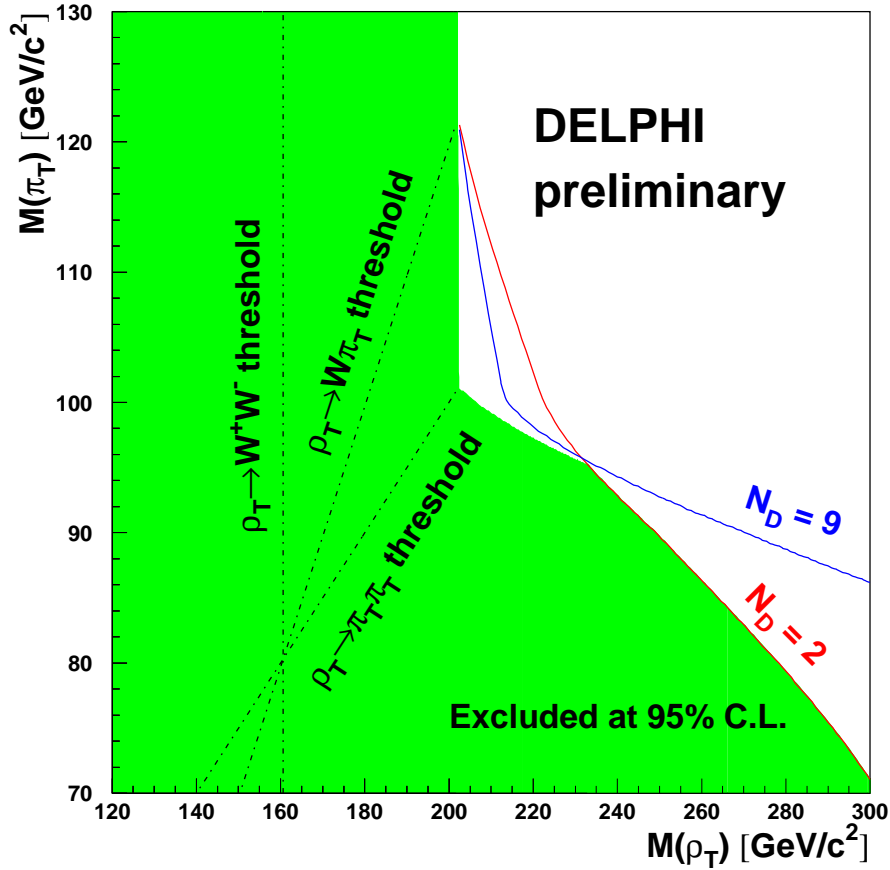}
\vskip3.0truecm
 \caption{\it
      The $M_{\tro}$--$M_{\tpi}$ region excluded at the 95\% CL by the DELPHI
      analysis of Ref.~[31].
    \label{fig10} }
\end{figure}
\begin{figure}[t]
 \vspace{9.0cm}
\includegraphics{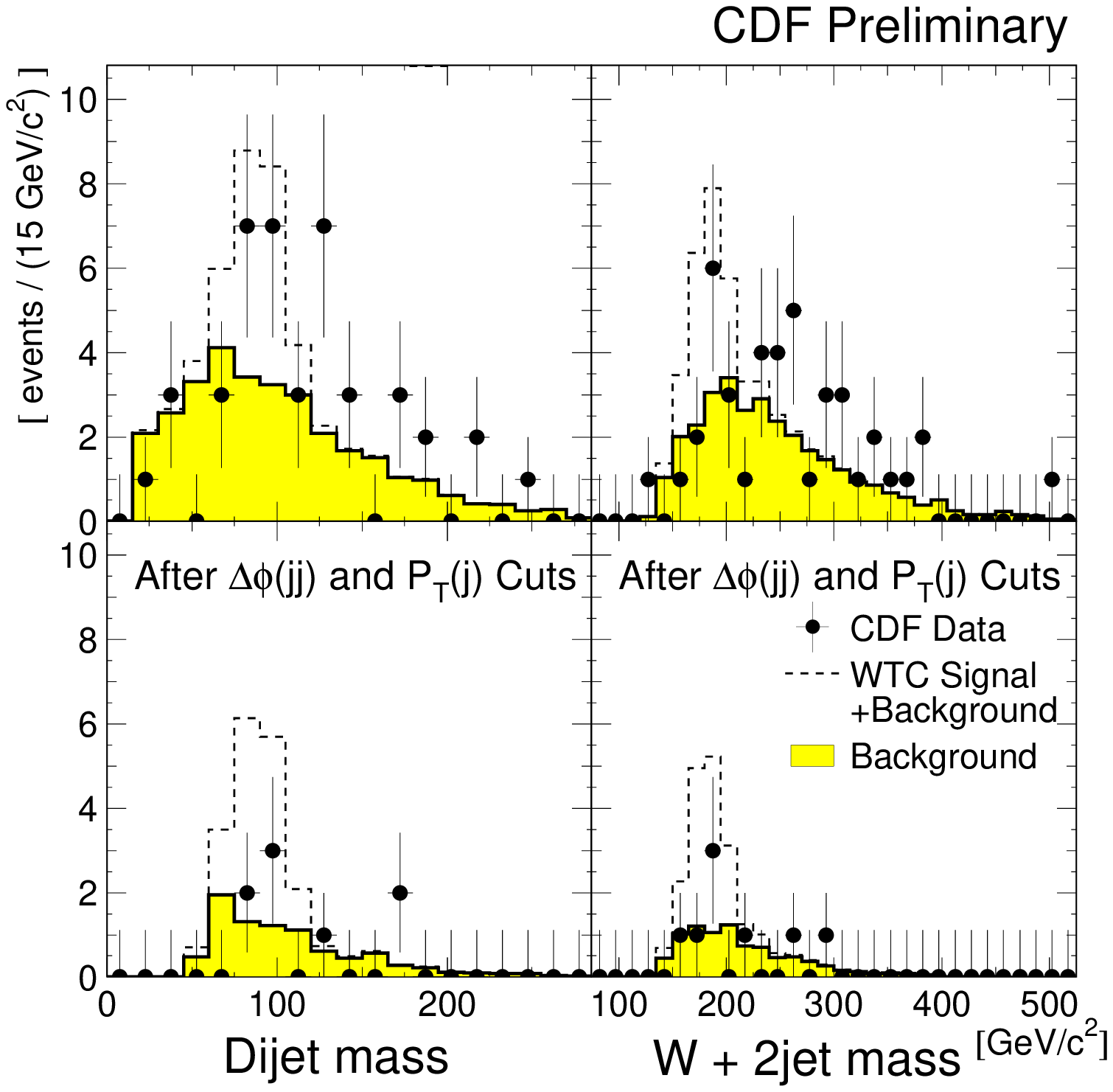}
\vskip2.3truecm
 \caption{\it
      Invariant mass of the dijet system and of the $W+2\ts\jet$ system for
      the $\ell+2\ts\jet$ mode; from Ref.~[32]. The mass combination shown is
      $M_{\tpi} =90\,\gev$ and $M_{\tro}=180\,\gev$.
    \label{fig11} }
\end{figure}
\begin{figure}[t]
 \vspace{9.0cm}
\includegraphics{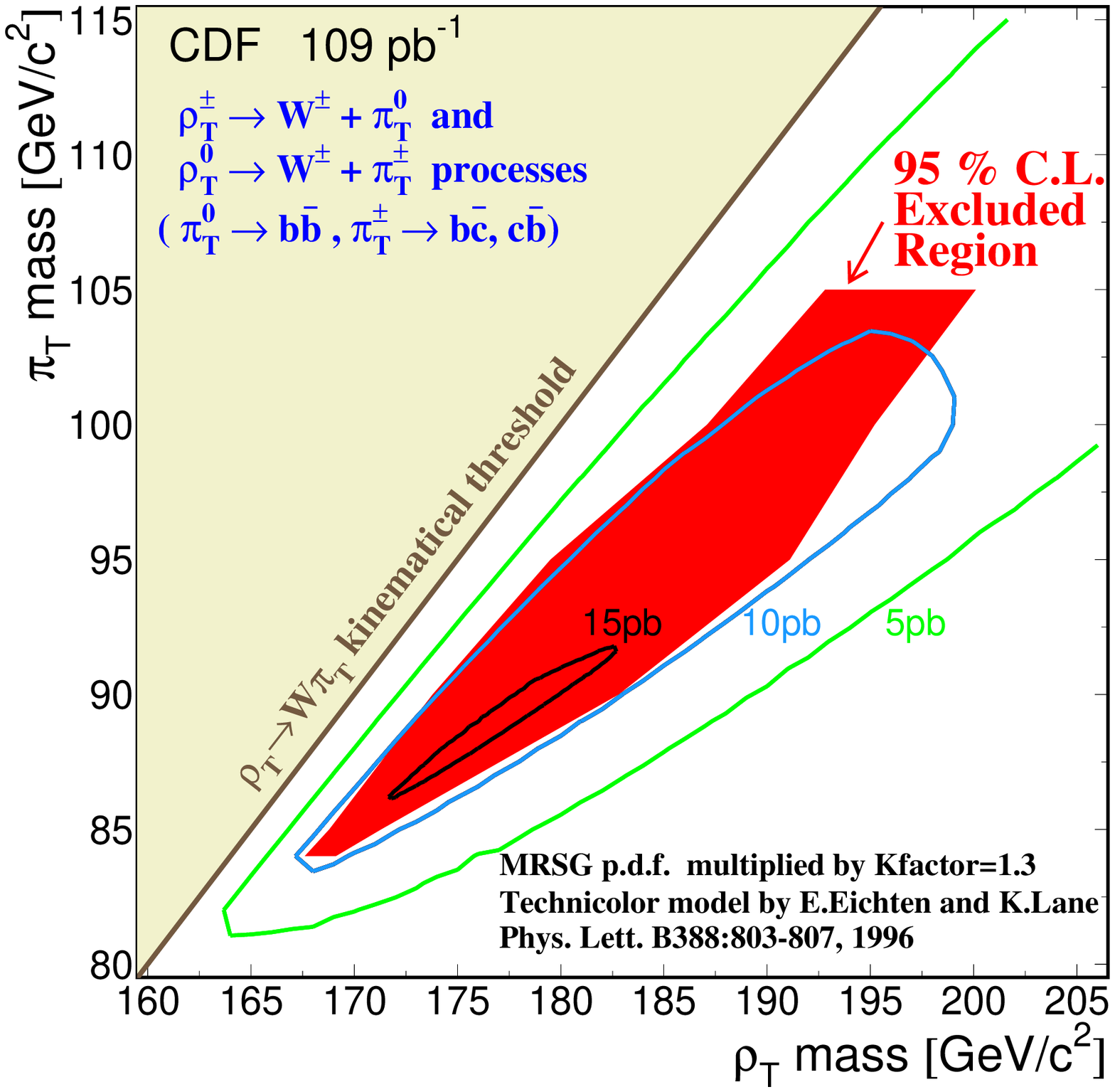}
\vskip2.3truecm
 \caption{\it
   Excluded region for the CDF search for $\tro \ra W^\pm \tpi$ in
   Ref.~[32]. See that reference for an explanation of the 5, 10, 15~pb
   contours.
    \label{fig12} }
\end{figure}
\section*{3. Today}
\subsection*{\underbar{Theoretical Issues: Walking Technicolor and
    Topcolor--Assisted Technicolor}}

{\begin{table}{
\centering
\begin{tabular}{|c|c|c|}
\hline
$V_T$ Decay Mode& 
$V(V_T \ra G\tpi) \times M_V/e$ & 
$A(V_T \ra G\tpi) \times M_A/e$  
\\
\hline\hline
$\tom \ra \gamma \tpiz$& $\cos\chi$ & 0 \\
$\ts\ts\ts\quad \ra \gamma \tpipr$ & $(Q_U + Q_D)\ts \cos\chi'$ & 0 \\ 
$\qquad \ra Z^0 \tpiz$ & $\cos\chi\cot 2\thw$ & 0 \\ 
$\ts\qquad \ra Z^0 \tpipr$ & $-(Q_U+Q_D)\ts \cos\chi'\tan \thw$ & 0 \\ 
$\ts\ts\ts\ts\qquad \ra W^\pm \tpimp$ & $\cos\chi/(2\sin\thw)$ & 0 \\ 
\hline
$\troz \ra \gamma \tpiz$ & $(Q_U + Q_D)\ts \cos\chi$ & 0 \\
$\ts\ts\ts\quad \ra \gamma \tpipr$ & $\cos\chi'$ & 0 \\
$\qquad \ra Z^0 \tpiz$ & $-(Q_U+Q_D)\ts \cos\chi \tan \thw$ & 0 \\
$\ts\qquad \ra Z^0 \tpipr$ & $\cos\chi'\ts \cot 2\thw$ & 0 \\
$\ts\ts\ts\ts\qquad \ra W^\pm \tpimp$ & 0 & $-\cos\chi/(2\sin\thw)$ \\
\hline
$\tropm \ra \gamma \tpipm$ & $(Q_U + Q_D)\ts \cos\chi$ & 0 \\ 
$\qquad \ra Z^0 \tpipm$ & $-(Q_U+Q_D)\ts \cos\chi \tan \thw$ & $\cos\chi
\ts /\sin 2\thw$ \\  
$\ts\ts\ts\qquad \ra W^\pm \tpiz$ & 0 & $\cos\chi/(2\sin\thw)$ \\ 
$\ts\ts\ts\qquad \ra W^\pm \tpipr$ & $\cos\chi'/(2\sin\thw)$ & 0 \\
\hline\hline
\end{tabular}}
\medskip
\caption{{\it Relative vector and axial vector amplitudes for $V_T \ra G
    \tpi$ with $V_T = \tro,\tom$ and $G$ a transverse electroweak boson,
  $\gamma,Z^0,W^\pm$; from Ref.~[26].}}
%
\end{table}}

\begin{figure}[t]
 \vspace{9.0cm}
\includegraphics{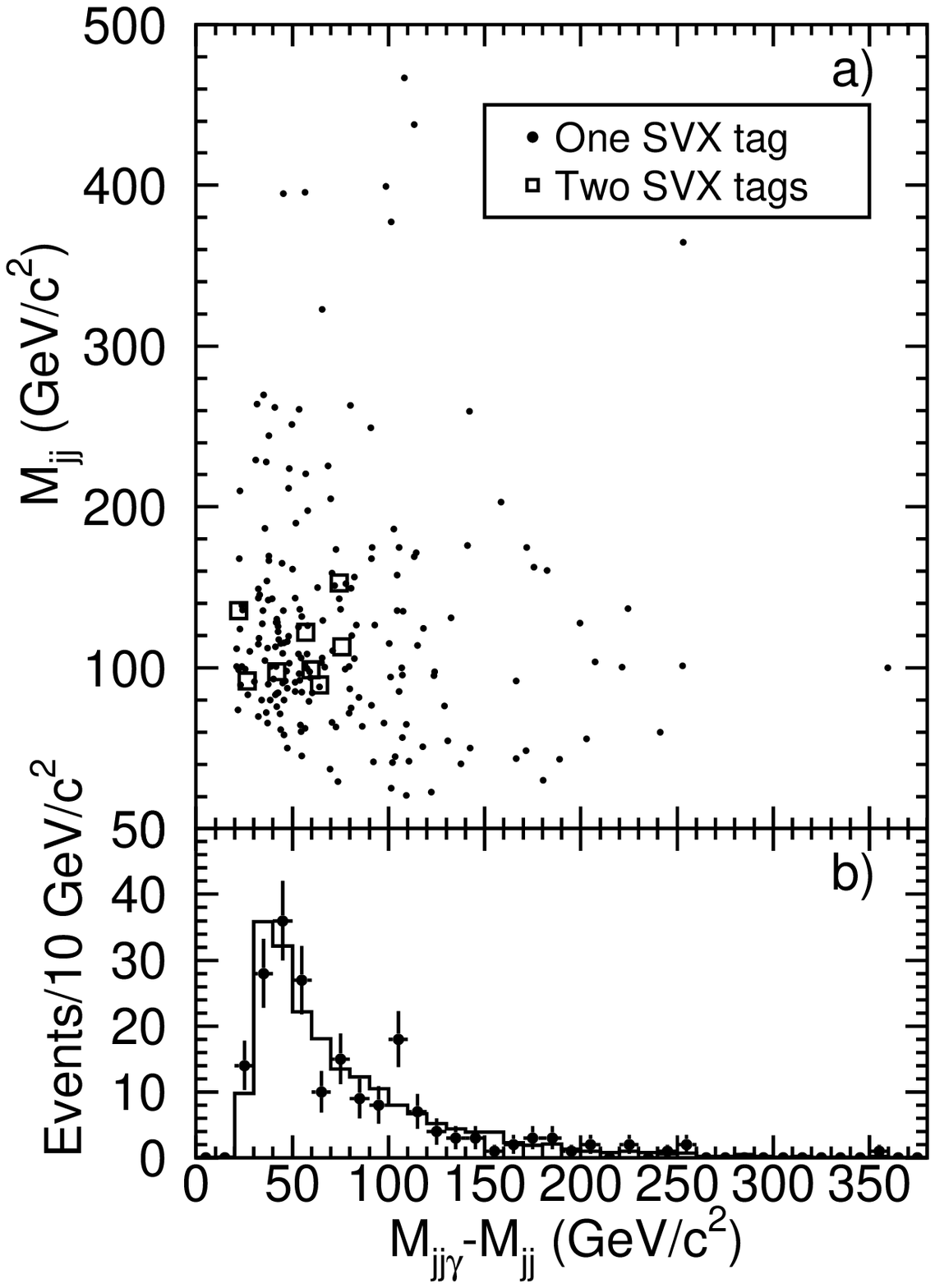}
\vskip2.0truecm
 \caption{\it
      (a) The distribution of $M_{jj}$ vs. $M_{jj\gamma} - M_{jj}$ for events
      with a photon, $b$--tagged jet and a second jet. (b) Projection of this
      data in $M_{jj\gamma} - M_{jj}$; from Ref.~[33]
    \label{fig13} }
\end{figure}

The FCNC and STU difficulties of technicolor have a common cause: the
assumption that technicolor is just a scaled--up version of QCD. In a
QCD--like technicolor theory, asymptotic freedom sets in quickly above
$\LTC$, $\gamma_m \ll 1$, and $\condetc \simeq \condtc$. The conclusion that
fermion and technipion masses are one or more orders of magnitude too small
then follows from the requirement in Eq.~(\ref{eq:fcnc}) that $\METC >
100\,\tev$. Scaling from QCD also means that the technihadron spectrum is
just a magnified image of the QCD--hadron spectrum, hence that $S$ is too
large for all technicolor models except, possibly, the minimal one--doublet
model with $N_{TC} \simle 4$. A solution to these difficulties in a
technicolor theory lies in gauge dynamics that are distinctly not
QCD--like. A technicolor theory in which the gauge coupling evolves slowly,
or ``walks'', is the only promising example of this.\cite{wtc}

In walking technicolor, the gauge coupling $\atc(\mu)$ remains close to its
critical value, the one required for spontaneous chiral symmetry breaking,
for scales $\LTC < \mu \simle \METC$. This implies that the anomalous
dimension $\gamma_m(\mu) \simeq 1$ in Eq.~(\ref{eq:condrenorm}), enhancing
the condensate $\condetc$ by a factor of 100 or more. This yields quark
masses up to a few GeV and reasonably large technipion masses despite the
very large ETC mass scale. This is still not enough to account for the top
mass; more on that momentarily.

Another consequence of the walking $\atc$ is that the spectrum of
technihadrons, especially $\tro$ and $\tom$, cannot be
QCD--like.\cite{kltasi,edrta} If it were, the integrals appearing in Weinberg's
spectral function sum rules~\cite{sfsr} would converge much more rapidly than
they do in a walking theory. As mentioned, there must be a tower of $\tro$
and $\tom$ extending up to $\METC$. How these affect the spectral integrals
that define $S$ is unknown. Another issue that may affect $S$ is that it is
usually defined assuming that the new physics appears at energies well above
$M_{W,Z}$. We shall see below that, on the contrary, walking technicolor
suggests that there are $\tpi $ and $\tro$ starting near or not far above
$100\,\gev$.

The large value of $\METC$ makes it difficult if not impossible to explain
the top mass by the conventional ETC mechanism, Eq.~({\ref{eq:qlmass}). The
most plausible dynamical explanation assumes another gauge interaction that
is strong near 1~TeV. This interaction, called topcolor, is like technicolor
for the third generation, but it must be more complicated to avoid making
$m_b = m_t$. The variant we describe here is called topcolor--assisted
technicolor (TC2).

In TC2, as in many top-condensate models of electroweak symmetry
breaking,\cite{topcondref} almost all of the top quark mass arises from the
strong topcolor interaction.\cite{topcref} To maintain electroweak symmetry
between (left-handed) top and bottom quarks and yet not generate $m_b \simeq
m_t$, the topcolor gauge group under which $(t,b)$ transform is usually taken
to be $SU(3)\otimes U(1)$. The $U(1)$ acts differently on $t_R$ and $b_R$
and, so, provides the difference that causes only top quarks to
condense. Then, in order that topcolor interactions be natural---i.e., that
their energy scale not be far above $m_t$---without introducing large weak
isospin violation, it is necessary that electroweak symmetry breaking is
still mainly due to technicolor interactions.\cite{tctwohill} Extended
technicolor interactions are still needed in TC2 models to generate the
masses of light quarks and the bottom quark, to contribute a few~GeV to
$m_t$,~\footnote{Massless Goldstone ``top-pions'' arise from top-quark
condensation. This ETC contribution to $m_t$ is needed to give them a mass in
the range of 150--250~GeV.} and to give mass to technipions. The scale of ETC
interactions still must be hundreds of~TeV to suppress flavor-changing
neutral currents and, so, the technicolor coupling still must walk.

Walking technicolor requires that the beta--function $\beta(\atc)$ be near
zero for a large range of energy above $\LTC$. This requires many
technifermions in the fundamental representation of $\sutc$, or a few in
higher-dimensional representations, or both.\cite{multiklee} The TC2 models
that are consistent with the pattern of quark masses and the mixings between
the heavy and light generations also require many ($\sim 10$)
technidoublets.\cite{tctwoklee,tctwokl}

All this suggests that the technicolor scale is much lower than previously
thought. If the number $N$ of technidoublets is $\CO(10)$, then $\LTC \simeq
F_T = F_\pi/\sqrt{N} \simle 100\,\gev$. This sets the mass scale for the
lightest color--singlet technivector mesons, $M_{\tro} \simeq M_{\tom} \simeq
2\LTC \simle 200\,\gev$. The lightest color--octet $\tro$, formed from
color--triplet technifermions (which are needed in TC2) will be heavier,
starting, perhaps, at 400--500~GeV. There are $4N^2 - 4$ technipions in
addition to $W^\pm_L$ and $Z^0_L$. The color--singlet $\tpi$ may have masses
as low as 100~GeV. In short, unlike the situation in minimal technicolor,
these technihadrons may be within reach of Tevatron Run~II. They are
certainly accessible at the LHC. Color--singlet $\tro$ and $\tom$ may even be
detected at LEP200. If the NLC or a muon collider is built, it will be able
carry out precision studies of color--singlet technihadrons. We turn now to
the signatures of this ``low--scale
technicolor''~\cite{elw,tcsm_singlet,tcsm_octet}:~\footnote{Many of these
signatures are now encoded in {\sc Pythia}.\cite{pythia}} How are the $\tro$,
$\tom$, and $\tpi$ produced, and how do they decay?

Color--singlet $\tro$ and $\tom$ are produced in the $s$--channel of $\ol q
q$ and $e^+ e^-$ annihilation. Color--octet $\troct$ are produced in $\ol q
q$ and $gg$ collisions. In QCD--like technicolor, they decay mainly to
two or more technipions, with $\troct$ decaying to color--octet and
color--triplet (leptoquark) pairs. Walking technicolor dramatically changes
this expectation.

In the extreme walking limit, $\condetc \simeq (\METC/\LTC) \condtc$, the
technipions have masses of order $\LTC$, and they are not pseudoGoldstone
bosons at all. Though this extreme limit is theoretically problematic because
it is exactly scale--invariant, it is clear that walking TC enhances $\tpi$
masses significantly more than it does the $\tro$ and $\tom$ masses. Thus, it
is likely that $M_{\tpi} \simge \half M_{\tro,\tom}$ and, so, the nominal
isospin--conserving decay channels $\tro \ra \tpi\tpi$ and $\tom \ra
\tpi\tpi\tpi$ are {\it closed}.\cite{multiklee} We discuss our expectations
first for the color--singlet sector, then for color--nonsinglets.

\subsection*{\underbar{Theory and Experiment for Color--Singlet
    Technihadrons}}

The flavor problem is hard whether it is attacked with extended technicolor
or any other weapon. After all these years nobody has a complete solution or
even a promising scenario. In the absence of an explicit ETC model, or any
other kind of model, we need experimental guidance. Experimentalists, in
turn, need guidance from theorists to constrain their schemes. Supersymmetry
has its MSSM. What follows is a description of the corresponding thing for
technicolor, in the sense that it defines a set of incisive experimental
tests in terms of a limited number of adjustable parameters. I call this the
``Technicolor Straw Man'' model (or TCSM). First, I'll outline it in the
color--singlet sector.

In the TCSM, we assume that we can consider {\it in isolation} the
lowest-lying bound states of the lightest technifermion doublet, $(T_U,
T_D)$. These technifermions are likely to be color singlets because,
otherwise, color-$SU(3)$ interactions would contribute significantly to their
hard (or current--algebra) mass.\cite{multiklrm} We shall assume that they
transform according to the fundamental representation of the technicolor
gauge group, $SU(\Ntc)$. Their electric charges are $Q_U$ and $Q_D =
Q_U-1$. The bound states in question are vector and pseudoscalar mesons. The
vectors include a spin--one isotriplet $\tro^{\pm,0}$ and an isosinglet
$\tom$. In topcolor--assisted technicolor, there is no need to invoke large
isospin--violating extended technicolor interactions to explain the
top--bottom splitting. Thus, techni--isospin can be, and likely must be, a
good approximate symmetry. Then, $\tro$ and $\tom$ will be mostly isovector
and isoscalar, respectively, and they will be nearly degenerate. Their
production in annihilation processes is described using vector meson
dominance and propagator matrices mixing them with $W^\pm$ and $\gamma$,
$Z^0$; see Ref.~[26], called TCSM--1 below. Again, mixing of these $\tro$ and
$\tom$ with their excitations is ignored in the TCSM.

The lightest pseudoscalar $\ol T T$ bound states, the technipions, also
comprise an isotriplet $\Pi_T^{\pm,0}$ and an isosinglet $\Pi_T^{0
\prime}$. However, these are not mass eigenstates; all color--singlet
isovector technipions have a $W_L$ component. To limit the number of
parameters in the TCSM, we make the simplifying assumption that the
isotriplets are simple two-state mixtures of the $W_L^\pm$, $Z_L^0$ and the
lightest mass-eigenstate pseudo-Goldstone technipions $\tpi^\pm, \tpiz$:
\be\label{eq:pistates}
 \vert\Pi_T\rangle = \sin\chi \ts \vert
W_L\rangle + \cos\chi \ts \vert\tpi\rangle\ts.
\ee
Here, $\sin\chi = F_T/F_\pi = 1/\sqrt{N} \ll 1$.

Similarly, $\vert\Pi_T^{0 \prime} \rangle = \cos\chipr \ts
\vert\tpipr\rangle\ + \cdots$, where $\chipr$ is another mixing angle and the
ellipsis refers to other technipions needed to eliminate the two-technigluon
anomaly from the $\Pi_T^{0 \prime}$ chiral current. It is unclear whether,
like $\troz$ and $\tom$, the neutral technipions $\tpiz$ and $\tpipr$ will be
degenerate as we have previously supposed.\cite{elw} On one hand, they both
contain the lightest $\ol T T$ as constituents. On the other, $\tpipr$ must
contain other, presumably heavier, technifermions as a consequence of anomaly
cancellation. We assume that $\tpiz$ and $\tpipr$ are nearly degenerate. If
this is true, and if their widths are roughly equal, there will be
appreciable $\tpiz$--$\tpipr$ mixing. Then, the lightest neutral technipions
will be ideally-mixed $\ol T_U T_U$ and $\ol T_D T_D$ bound states. In any
case, the technipions, assumed here to be lighter than $m_t + m_b$, are
expected to decay as follows: $\tpip \ra c \ol b$ or $c \ol s$ or even
$\tau^+ \nu_\tau$; $\tpiz \ra b \ol b$ and, perhaps $c \ol c$,
$\tau^+\tau^-$; and $\tpipr \ra gg$, $b \ol b$, $c \ol c$,
$\tau^+\tau^-$. This puts a premium on heavy--flavor identification in
collider experiments. However, this is only an educated guess and it is
possible that the mass--eigenstate neutral $\tpi$ have a sizable branching
ratio to gluon (or even light--quark) pairs.


For vanishing electroweak couplings $g,g'$, the $\tro$ and $\tom$ decay as
\bea\label{eq:vt_decays}
\tro &\ra& \Pi_T \Pi_T = \cos^2 \chi\ts (\tpi\tpi) + 2\sin\chi\ts\cos\chi
\ts (W_L\tpi) + \sin^2 \chi \ts (W_L W_L) \ts; \nn \\
\tom &\ra& \Pi_T \Pi_T \Pi_T = \cos^3 \chi \ts (\tpi\tpi\tpi) + \cdots \ts.
\eea
As noted above however, the all--$\tpi$ modes are likely to be closed. Thus,
major decay modes of the $\tro$ will be $W_L\tpi$ or, if $M_{\tro} \simle
180\,\gev$ (a possibility we regard as unlikely, if not already eliminated by
LEP data), $W_L W_L$. The $W^\pm_L\tpi^{\mp, 0}$ and $Z^0_L\tpi^\pm$ decays
of $\tro$ have striking signatures in any collider. Only at LEP is it now
possible to detect $\troz \ra W^+W^-$ above background. If $M_{\tom} <
250\,\gev$, all the $\tom \ra \Pi_T \Pi_T \Pi_T$ modes are closed. In all
cases, the $\tro$ and $\tom$ are very narrow, $\Gamma(\tom) \simle
\Gamma(\tro) \simle 1\,\gev$, because of the smallness of $\sin\chi$ and the
limited phase space. Therefore, we must consider other decay modes. These are
electroweak, suppressed by powers of $\alpha$, but not by phase space.

The decays $\tro, \tom \ra G \tpi$, where $G$ is a transversely polarized
electroweak gauge boson, and $\tro, \tom \ra \ol f f$ were calculated in
TCSM--1. The $G \tpi$ modes have rates of $\CO(\alpha)$, while the fermion
mode $\ol f f$ rates are $\CO(\alpha^2)$. The $\Gamma(\tro, \tom \ra G \tpi)$
are suppressed by $1/M^2_V$ or $1/M^2_A$, depending on whether the vector or
axial vector part of the electroweak current is involved in the decay. Here,
$M_{V,A}$ are masses of order $\LTC$ occuring in the dimension--5 operators
for these decays. We usually take them equal and vary them from 100 to
400~GeV. For smaller values of $M_{V,A}$, these modes, especially the
$\gamma\tpi$ ones, are as important as the $W_L\tpi$ modes.  For larger
$M_{V,A}$ and $|Q_U + Q_D| \simge 1$, the $\ol f f$ decay modes may become
competitive. Table~1 lists the relative strengths of the decay amplitudes for
the $\tro, \tom \ra G \tpi$ processes. Figure~4 gives a sense of the
$M_{V,A}$ dependence of the total decay rates of $\tro$ and $\tom$ for
$M_{\tro} = 210\,\gev$, $M_{\tom} = 200$--$220\,\gev$, $M_{\tpi} =
110\,\gev$, and $Q_U = Q_D + 1 = 4/3$. Figure~5 shows the rates for $Q_U =
-Q_D = 1/2$. These and all subsequent calculations assume that $\Ntc = 4$ and
$\sin\chi=\sin\chi'=1/3$. Experimental analyses quoted below use the same
defaults and (usually) $Q_U = Q_D + 1 = 4/3$.

Figures~6 and~7 show the cross sections in $\ol p p$ collisions at
$\ecm=2\,\tev$ for production of $\gamma\tpi$ and for $W\tpi$, $Z\tpi$ and
$\tpi\tpi$ as a function of $M_V = M_A$. Figure~8 shows the $e^+e^-$ rate for
$M_V = 100\,\gev$.\cite{tcsm_singlet} In these three graphs, $Q_U = Q_D + 1 =
4/3$. The production rates in these figures, all in the picobarn range, are
typical for the Tevatron for $M_{\tro,\tom} \simle 250\,\gev$ and $M_{\tpi}
\simle 150\,\gev$. Thus, Run~II will probe a significant portion of
low--scale technicolor parameter space.

\begin{figure}[t]
\vspace{9.0cm}
\includegraphics{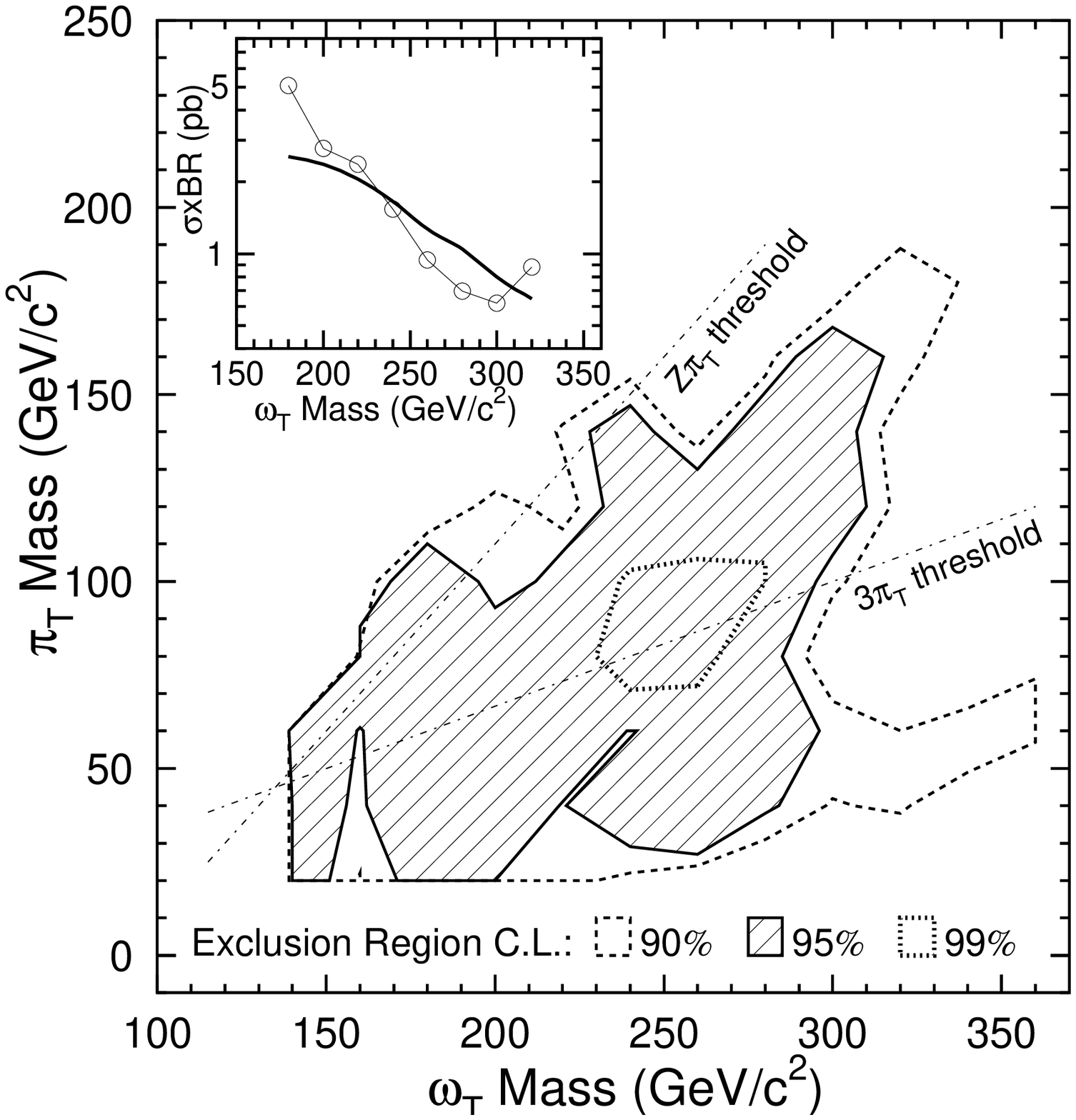}
\vskip2.5truecm
\caption{\it
  The 90\%, 95\% and 99\% CL exclusion regions for the CDF search for $\tom
  \ra \gamma\tpi$ in Ref.~[33]. The inset shows the limit on $\sigma B$ for
  $M_{\tpi} = 120\,\gev$. The circles represent the limit and the solid line
  the prediction from the second paper in Ref.~[25].
  \label{fig14} }
\end{figure}
%
%

%
\begin{figure}[t]
\vspace{9.0cm}
\includegraphics{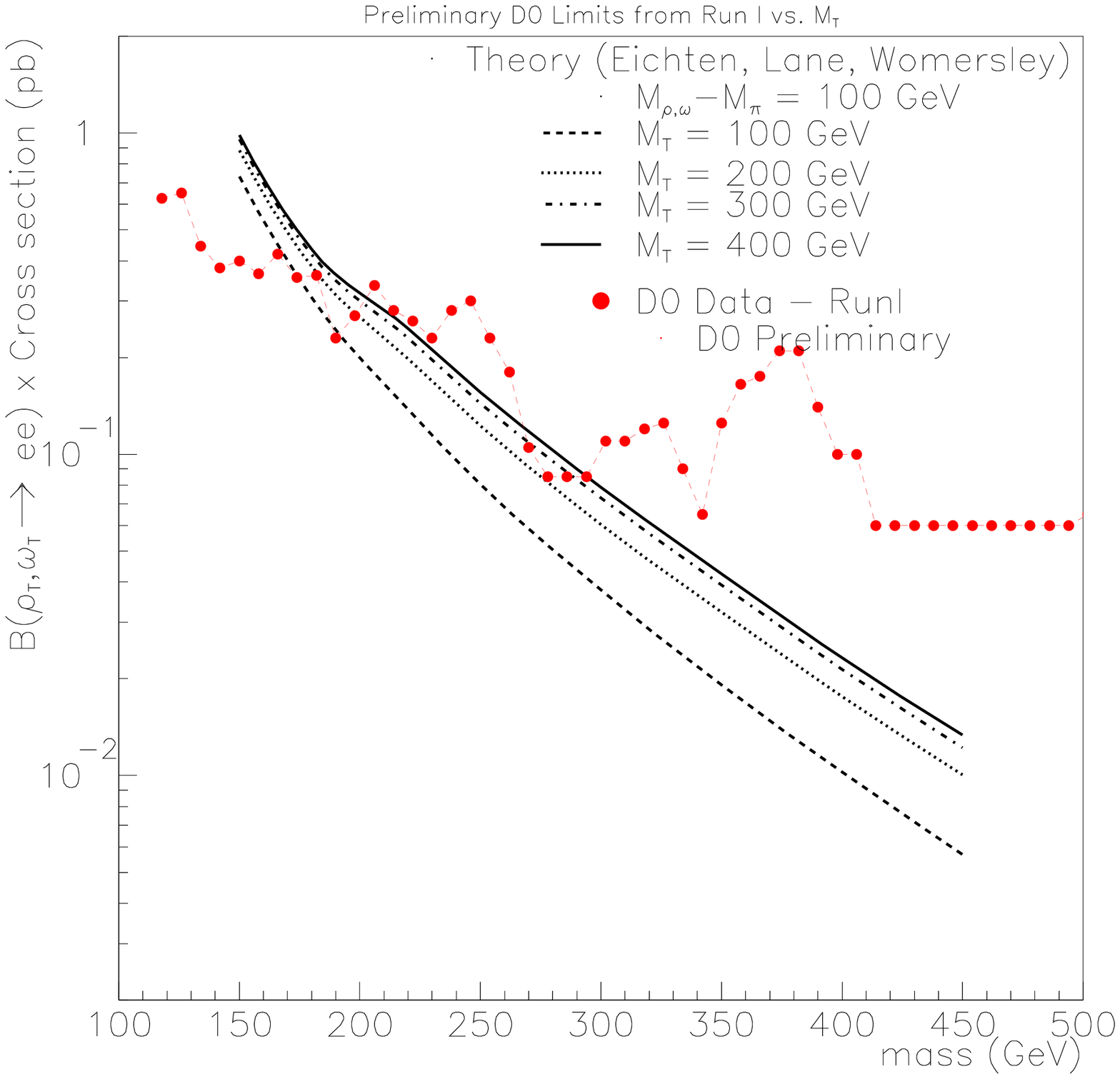}
\vskip1.5truecm
\caption{\it
  Excluded regions for the D\O\ search for $\troz,\tom \ra e^+e^-$; from
  Ref.~[34].
\label{fig15} }
\end{figure}
\begin{figure}[t]
\vspace{9.0cm}
\includegraphics{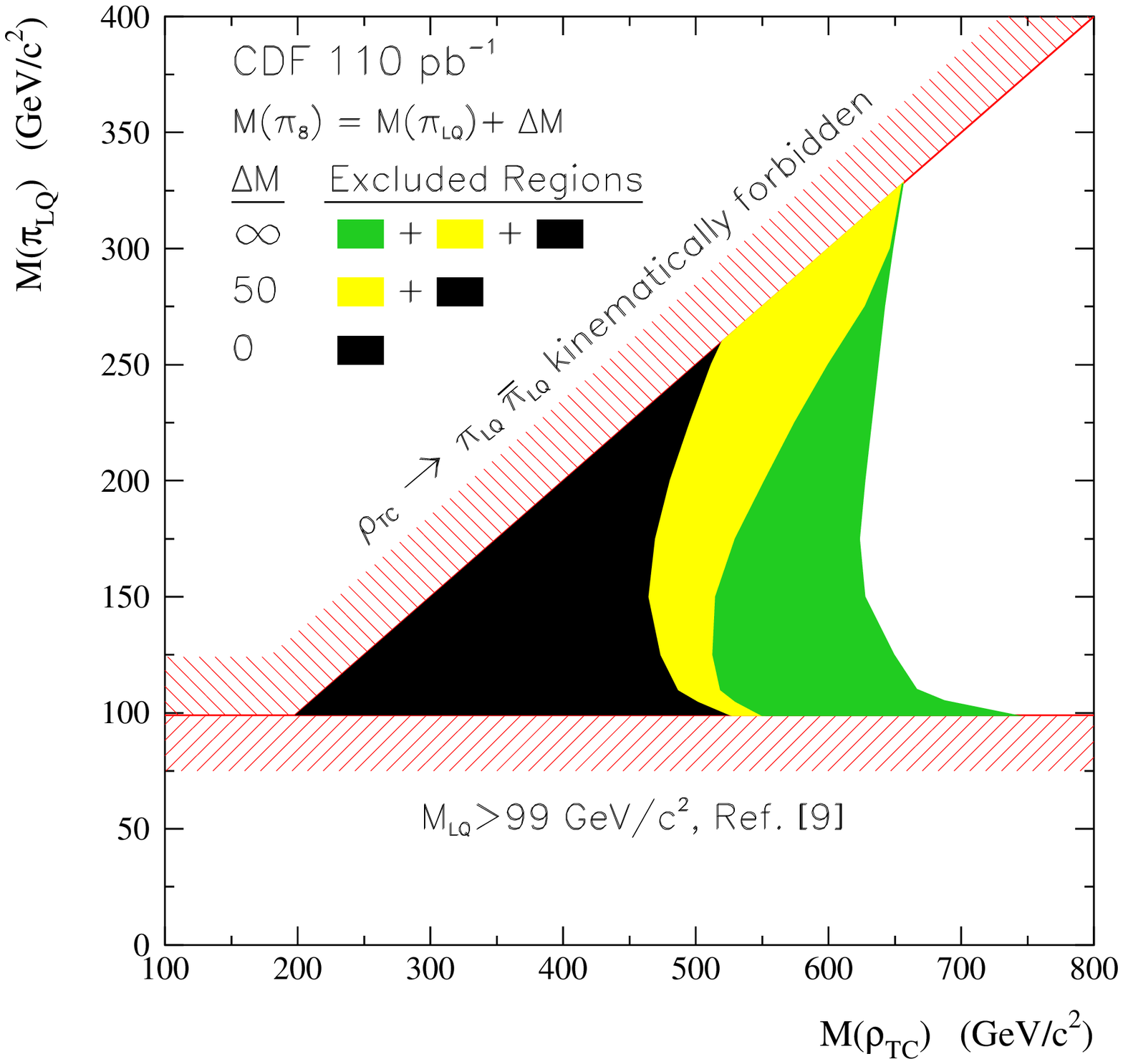}
\vskip2.8truecm
\caption{\it
  The 95\% CL exclusion regions for various $M_{\octpi} - M_{\tpilq}$ from a
  CDF search for $\troct \ra \tpilq\tpiql \ra \tau^+\tau^-\ts\jet\ts\jet$;
  from Ref.~[36].
\label{fig16} }
\end{figure}
\begin{figure}[t]
 \vspace{9.0cm}
\includegraphics{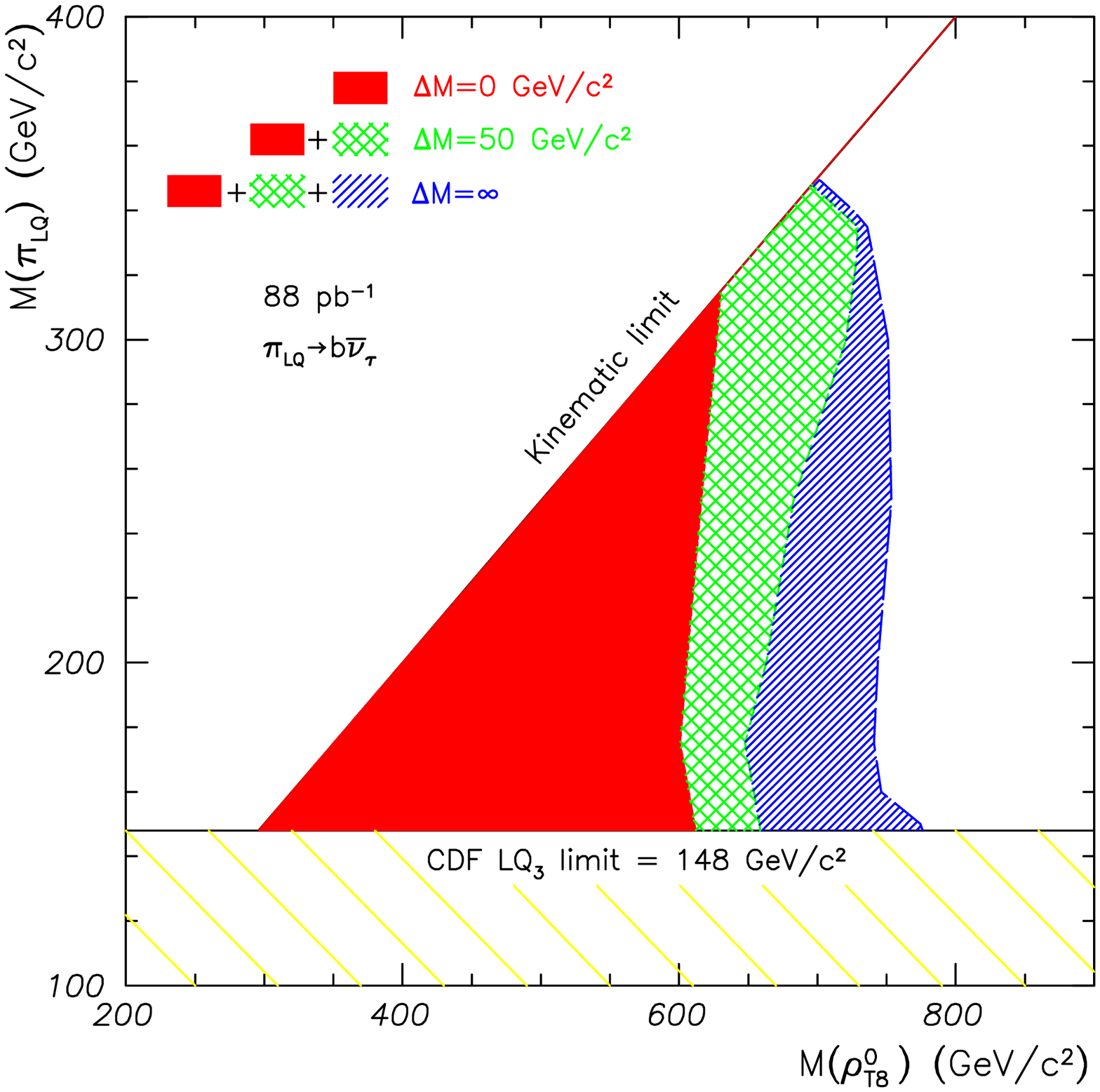}
\vskip2.7truecm
 \caption{\it
   The 95\% CL exclusion regions for various $M_{\octpi} - M_{\tpilq}$ from a
  CDF search for $\troct \ra \tpilq\tpiql \ra \ol b b \nu\nu$; from Ref.~[37].
    \label{fig17} }
\end{figure}
\begin{figure}[t]
 \vspace{9.0cm}
\includegraphics{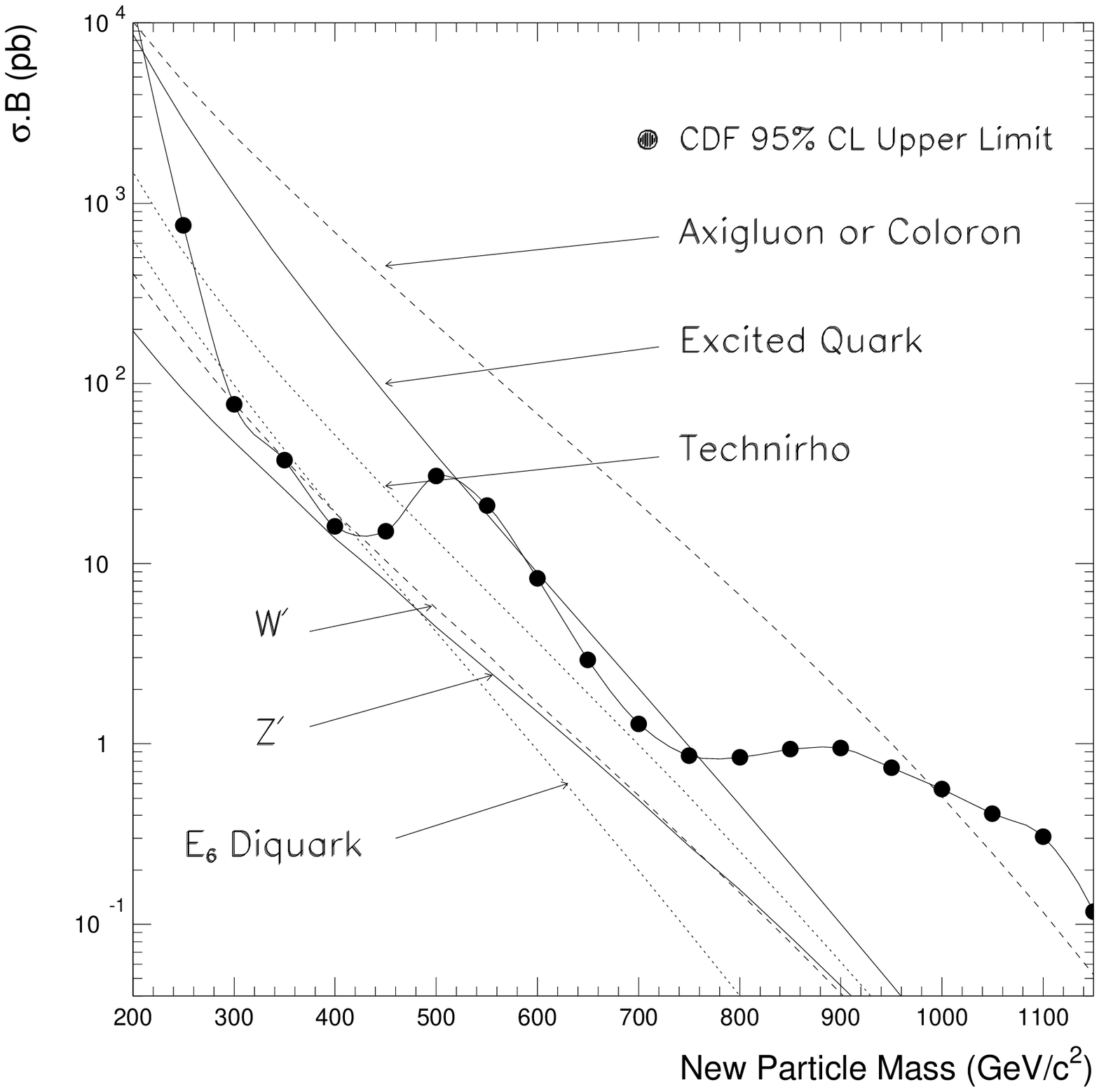}
\vskip1.8truecm
 \caption{\it
      The 95\% exclusions for a CDF search for $\troct \ra \jet\ts\ts\jet$ and
      other narrow dijet resonances; from Ref.~[38].
    \label{fig18} }
\end{figure}
\begin{figure}[t]
\vspace{9.0cm}
\includegraphics{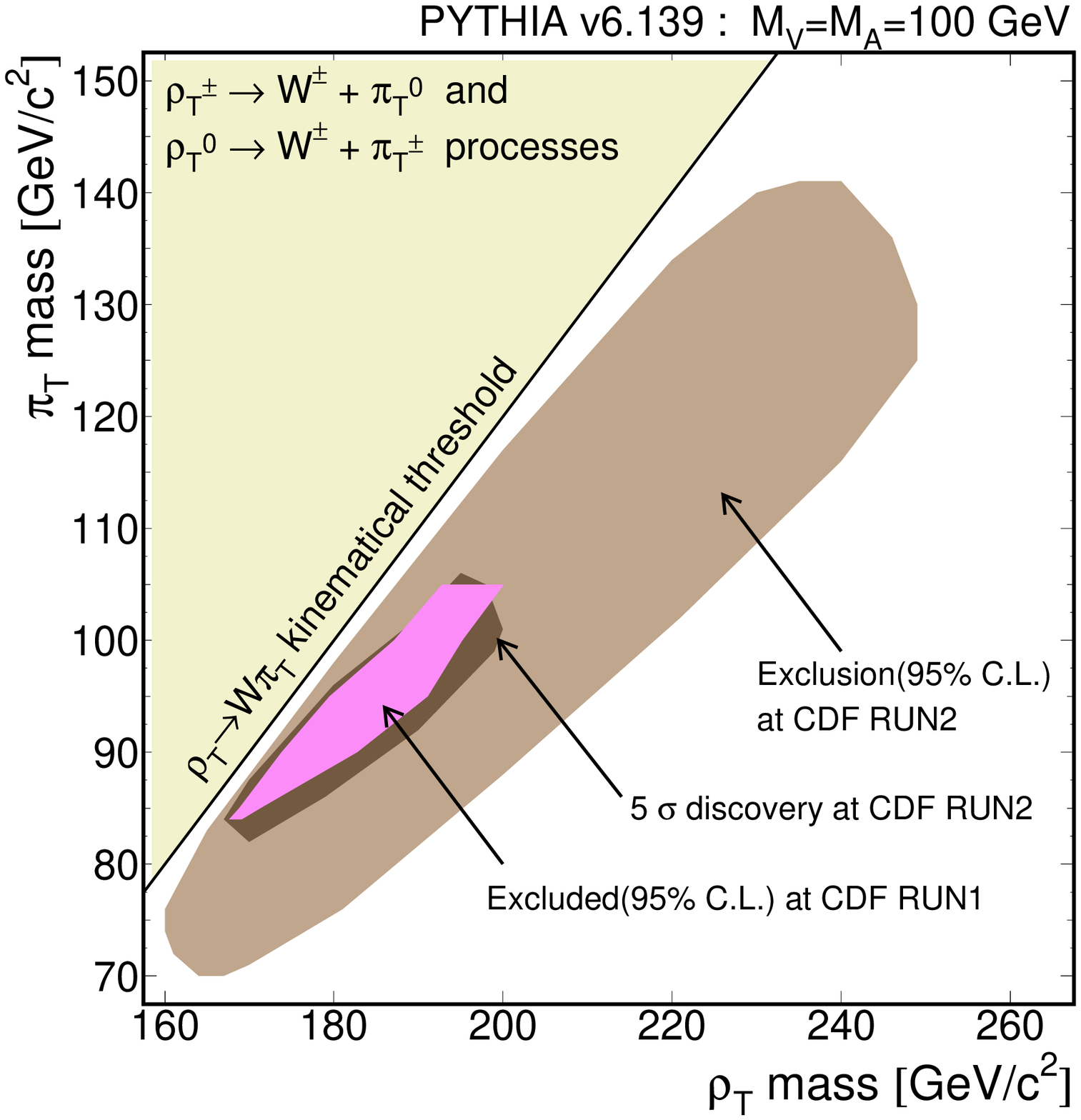}
\vskip2.7truecm
\caption{\it
  Reach of the CDF detector in Tevatron Run~IIa for $\tro \ra W^\pm \tpi$
  with $M_V = M_A = 100\,\gev$; from Ref.~[39].
\label{fig19} }
\end{figure}

\begin{figure}[t]
\vspace{9.0cm}
\includegraphics{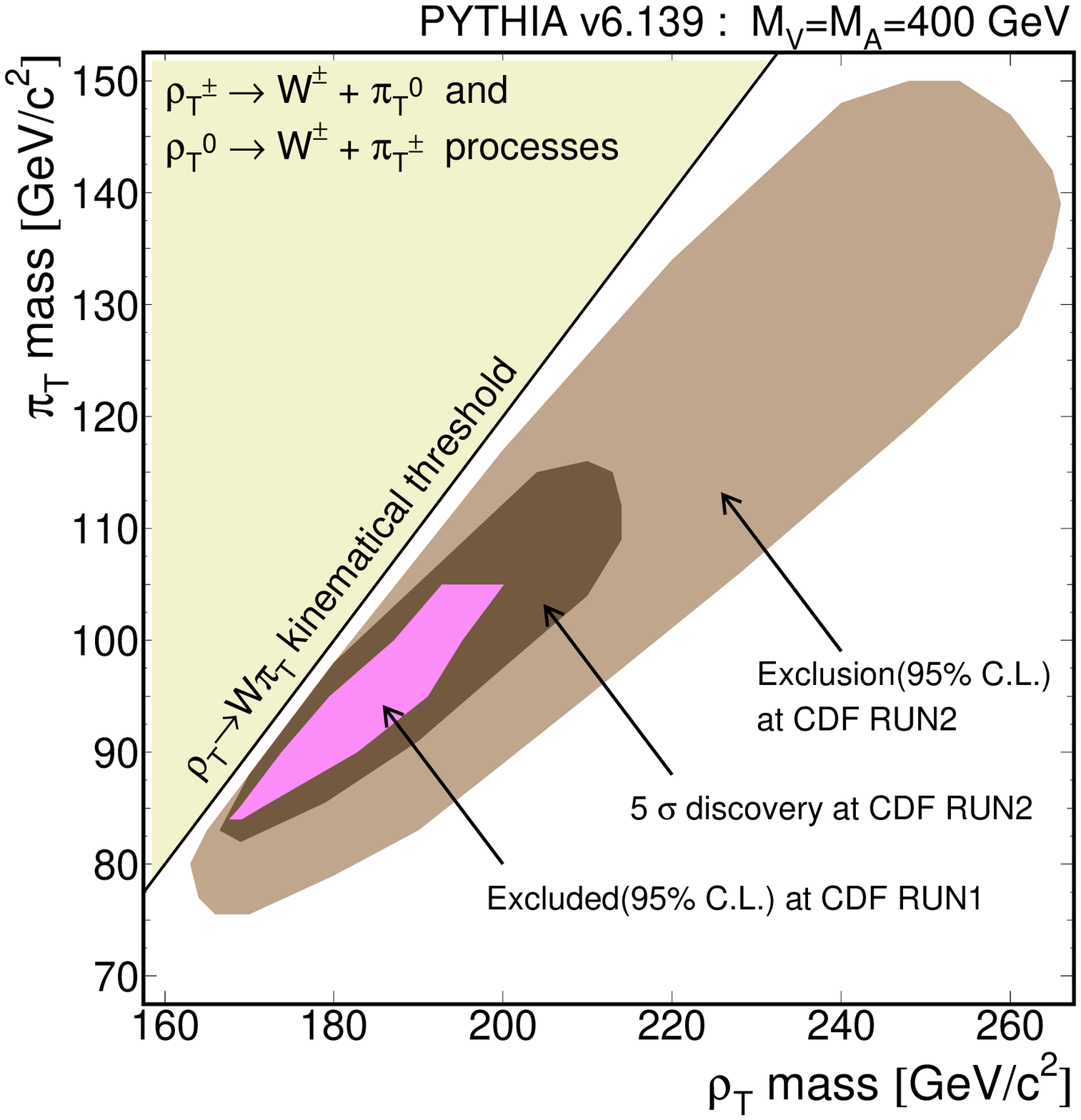}
\vskip2.7truecm
\caption{\it
  Reach of the CDF detector in Tevatron Run~IIa for $\tro \ra W^\pm \tpi$
  with $M_V = M_A = 400\,\gev$; from Ref.~[39].
\label{fig20} }
\end{figure}

\begin{figure}[t]
 \vspace{9.0cm}
\includegraphics{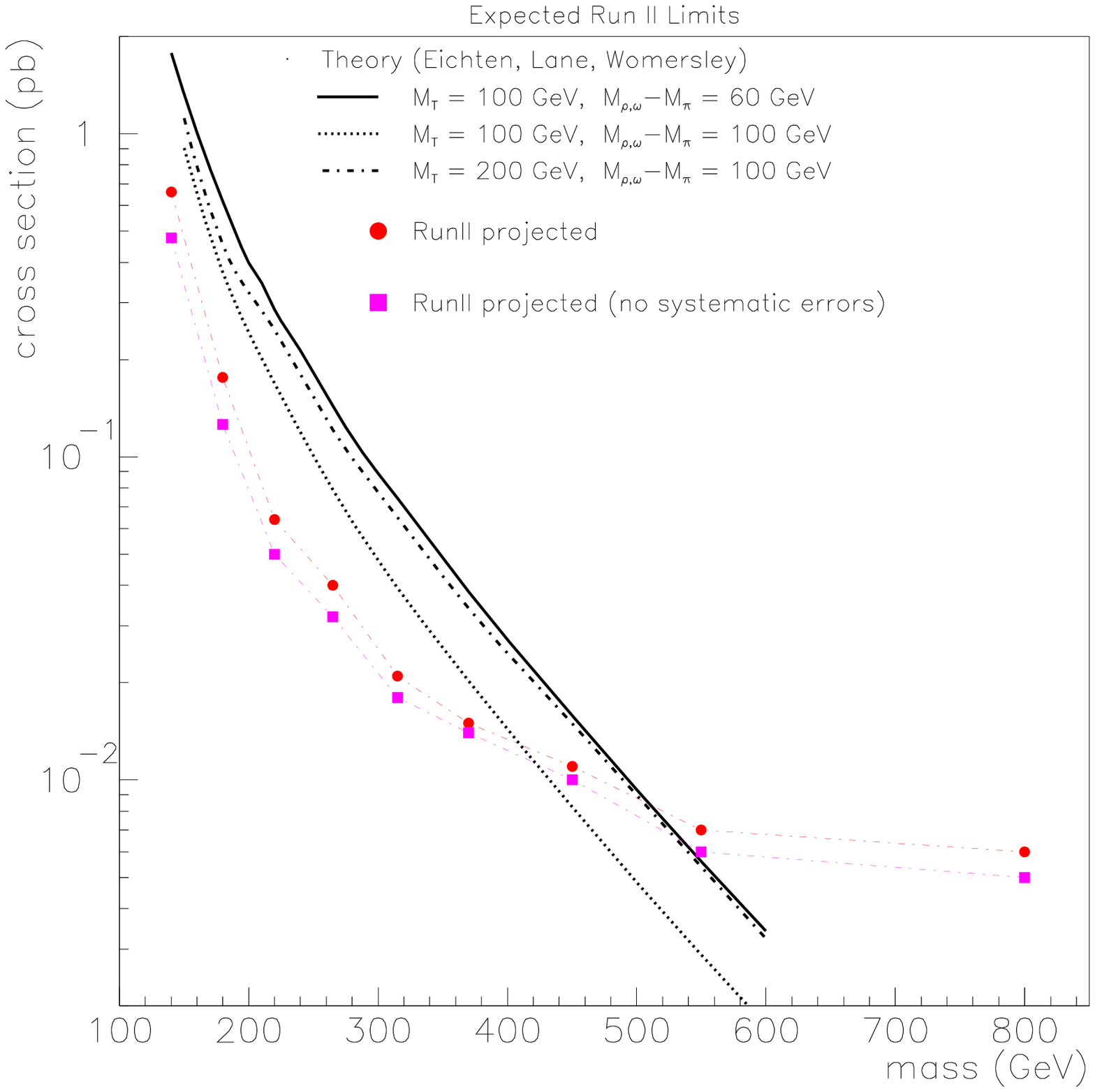}
\vskip2.0truecm
 \caption{\it
      Reach of the D\O\ detector in Tevatron Run~IIa for $\troz,\tom \ra
      e^+e^-$; from Ref.~[40].
    \label{fig21} }
\end{figure}
\begin{figure}[t]
 \vspace{9.0cm}
\includegraphics{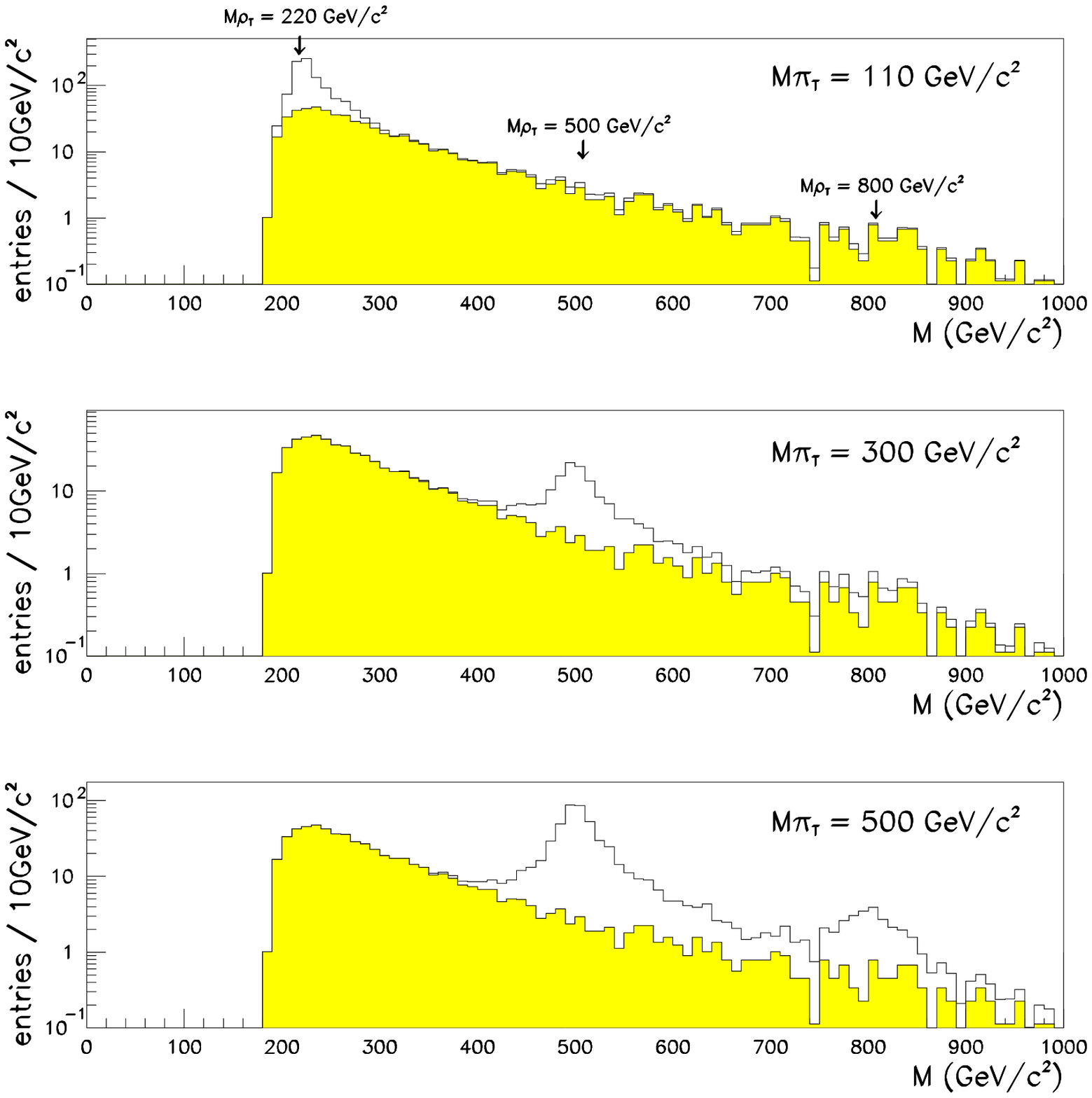}
\vskip5.0truecm
 \caption{\it
      Simulated event and background rates in the ATLAS detector for $\tropm
      \ra W^\pm Z \ra \ell^\pm\nu_\ell \ell^+\ell^-$ for various $M_{\tro}$
      and $M_{\tpi}$; from Ref.~[41].
    \label{fig22} }
\end{figure}

Turning to the recent searches for color--singlet technihadrons, we start
with a study by the L3~Collaboration at LEP.\cite{L3} This search is based on
$176\,\ipb$ of data taken an average energy of $189\,\gev$. The L3 analysis
used TCSM--1 and it studied the channels $e^+e^- \ra \troz \ra W^+W^-$;
$W^\pm_L \tpimp \ra \ell\nu_\ell bc$; $\tpip\tpim \ra c\ol b \ts b \ol c$;
and $\gamma \tpiz \ra \gamma b \ol b$. The TC--scale masses were fixed at
$M_V = M_A = 200\,\gev$ and the technifermion charges ranged over $Q_U + Q_D
= 5/3,0,-1$. The resulting 95\% confidence limits in the
$M_{\tro}$--$M_{\tpi}$ plane are shown in Fig.~9. A similar study was also
carried out by the DELPHI Collaboration\cite{delphi} and its exclusion plot
is shown in Fig.~10. Note that LEP experiments can be sensitive to $\tro$
masses significantly above the $e^+e^-$ cm energy, $\ecm$. This is because
the $e^+e^-$ cross section on resonance is very large for the narrow $\tro$.

Since these LEP analyses were done, I have realized the cross section
formulae stated in TCSM--1 are inappropriate for $\ecm$ well below
$M_{\tro}$.  This is unimportant for the Tevatron and LHC, where the
production rate comes mainly from integrating over the resonance
pole. However, it may have a significant effect on limits derived from
$e^+e^-$ annihilation. This is especially true for the $W^+W^-$ channel,
which has a large standard model amplitude interfering with the TCSM
one.~\footnote{I thank F.~Richard for drawing my attention to this
shortcoming of TCSM--1. A correction will be issued soon.} Another example is
that limits on $M_{\tpi} $ approaching $\ecm/2$ should be derivable from
$e^+ e^- \ra \tpi^+\tpi^-$.

The CDF Collaboration has analyzed Tevatron Run~I data to search for the
processes signalled by a $W$ or photon plus two jets, one of which is
$b$--tagged:
\bea\label{eq:cdfsearches}
\ol q q \ra W^\pm,\gamma,Z^0 &\ra& \tropm \ra W^\pm_L \tpi \ra \ell^\pm
\nu_\ell \ts b + \jet \nn\\
        &\ra& \tropm,\troz,\tom \ra \gamma \tpi \ra
        \gamma \ts b + \jet \ts.
\eea
These analyses were carried out before the publication of TCSM--1 so they do
not include the $G\tpi$ and $\ol ff$ processes and corresponding branching
ratios. They will be included in analyses of Run~II data. Figure~11 shows data
for the $W\tpi$ search on top of a background and signal expected for default
parameters with $M_{\tro} = 180\,\gev$ and $M_{\tpi} = 90\,\gev$. The
topological cuts leading to the lower figure are described in the second
paper of Ref.~[25]. The region excluded at 95\% confidence level is shown in
Fig.~12.\cite{cdfwpi}

Figure~13 shows the invariant mass of the tagged and untagged jets and the
invariant mass difference $M(\gamma+b+\jet) - M(b+\jet)$ in a search for
$\tom,\tro \ra \gamma \tpi$.\cite{cdfgpi} The good resolution in this mass
difference is controlled mainly by that of the electromagnetic energy.  The
exclusion plot is shown in Fig.~14. It is amusing that the $\sim 2\sigma$
excesses in Figs.~11 and 13 correspond to nearly the same
$M_{\tro,\tom}\simeq 200\,\gev$ and $M_{\tpi}\simeq 100\,\gev$.

The D\O\ Collaboration has studied its Run~I Drell--Yan data to search for
$\tro, \tom \ra e^+e^-$.\cite{dzeroee} The data and the excluded region are
shown in Fig.~15 for $Q_U = Q_D + 1 = 4/3$, $M_V = 100$--$400\,\gev$ and
$M_{\tro} - M_{\tpi} = 100\,\gev$. Increasing $M_V$ (called $M_T$ in the
figure) and decreasing $M_{\tro} - M_{\tpi}$ both increase the branching
ratio for the $e^+e^-$ channel, the former because it decreases $\tro, \tom
\ra \gamma\tpi$, the latter because it decreases $\tro \ra W\tpi$. For the
parameters considered here, $M_{\tro} = M_{\tom} < 150$--$200\,\gev$ is
excluded at the 95\% CL.

\subsection*{\underbar{Color--Nonsinglet Technihadrons}}

We turn now to the color--nonsinglet sector. So far, most of the experimental
searches have been inspired by the phenomenology of a pre--TCSM, one--family
TC model containing a single doublet each of color--triplet techniquarks $Q =
(U,D)$ and of color--singlet technileptons $L =
(N,E)$.\cite{fs,ehlq,multiklrm}. Therefore, we defer the details of the TCSM
for the color--nonsinglet sector to the next section. Assuming that
techni--isospin is conserved, production of color--nonsinglet states is
assumed to proceed through the lightest isoscalar color--octet technirho,
$\troct$:
%
\bea\label{eq:nonsing}
\ol q q, \ts gg  \ra g \ra \troct &\ra& \octpi\octpi \nn\\
        &\ra& \tpilq\tpiql \nn\\
        &\ra& \ol q q, \ts gg \ts \jets \ts.
\eea
Here, $\octpi = \octpipm, \octpiz, \pi_{T8}^{0'}\equiv\eta_T$ are four
color--octet technipions that are expected to decay to heavy $\ol q q$ pairs;
$\tpilq$ are four color--triplet ``leptoquarks'' expected to decay to heavy
$\ol \ell q$ with the corresponding charges. If TC2 is invoked, the neutral
$\octpi$ decay to $\ol b b$ and, possibly, $gg$ as readily as to $\ol t t$.

The only $\troct \ra \tpi\tpi$ searches so far are by CDF for leptoquarks
$\tpied \ra \tau^+ b$ where the $b$ is not tagged~\cite{cdfed} and for
$\tpind \ra \nu b$, $\nu c$.\cite{cdfnd} These are based on $110\,\ipb$ and
$88\,\ipb$ of Run~I data, respectively. The exclusion plot for the
$\tau^+\tau^-$dijet signal is shown in Fig.~16 as a function of the
$\octpi$--$\tpilq$ mass difference. The theoretically likely case is that
this mass difference is about 50~GeV, implying a 95\% excluded region
extending over $200 \simle M_{\troct} \simeq 2 M_{\tpilq} \simle
500\,\gev$. Figure~17 shows the reach for $\tros \ra b \ol b \nu \ol\nu$ with
at least one $b$--jet tagged. Here the 95\% limits extends over $300 \simle
M_{\troct} \simeq 2 M_{\tpilq} \simle 600\,\gev$. The search for $\tpilq \ra
c \nu$ excludes a similar range. These limits are quite impressive. However,
it is not clear that they will remain so when complications of TC2 are taken
into account in the color--nonsinglet sector. These will be discussed in the
next section.

Given the walking technicolor enhancement of $\tpi$ masses, it is likely that
the $\troct \ra \tpi\tpi$ channels are closed. In that case, one seeks
$\troct \ra \jet\ts\jet$ in $b$--tagged and untagged jets. The results of a
CDF search for narrow dijet resonances in Run~I is shown in
Fig.~18.\cite{cdfjets} The region $260 < M_{\troct} < 460\,\gev$ is excluded
at the 95\% confidence level. This, too, is a stringent constraint, but its
applicability to TC2 models is uncertain.

\section*{4. Tomorrow}
\begin{figure}[t]
 \vspace{9.0cm}
\includegraphics{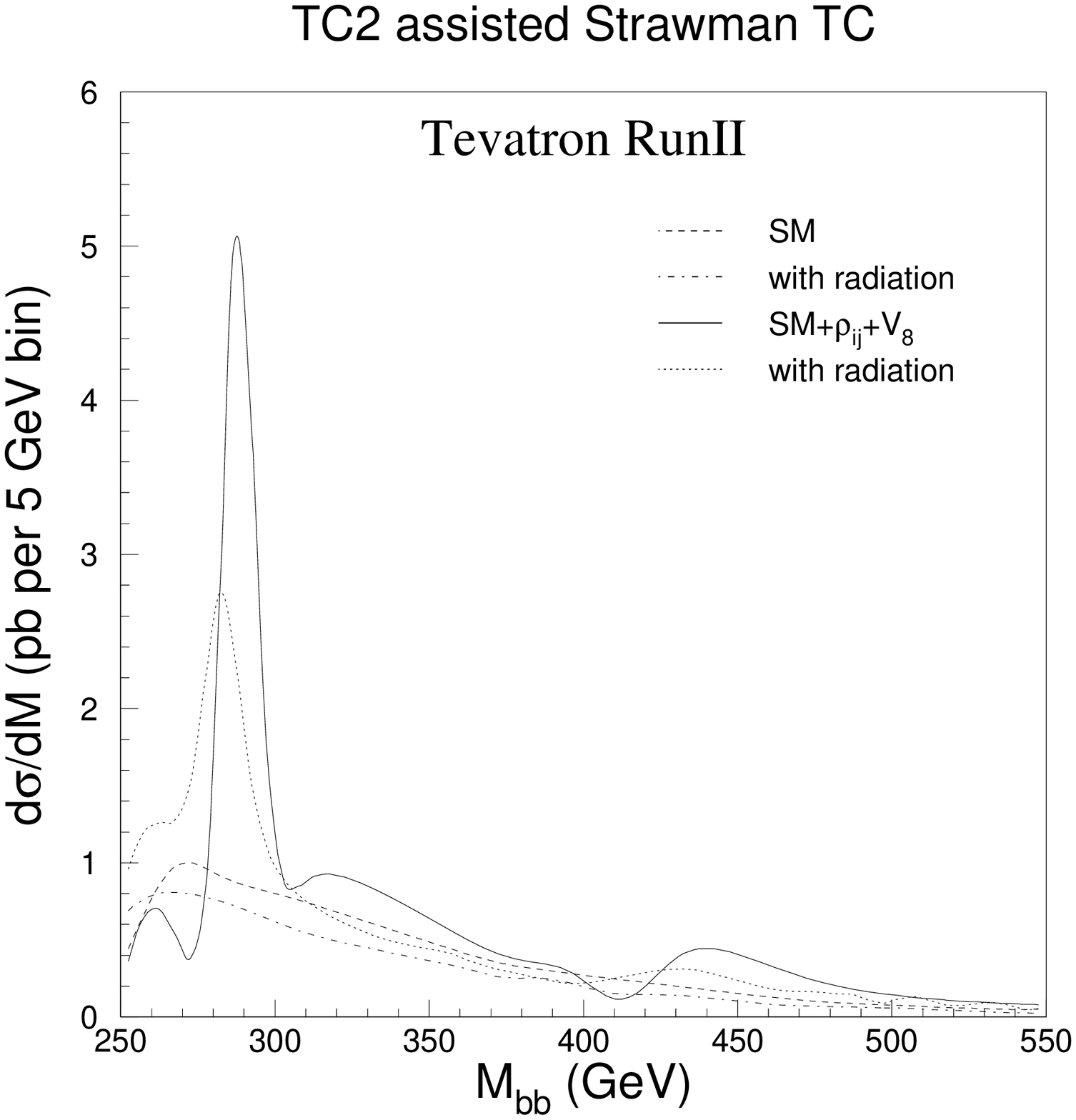}
\vskip1.8truecm
 \caption{\it
      Production of $\ol bb$ in $\ol pp$ collisions at $\ecm=2\,\tev$
      according to the TCSM model of Ref.~[27].
    \label{fig23} }
\end{figure}

\begin{figure}[t]
 \vspace{9.0cm}
\includegraphics{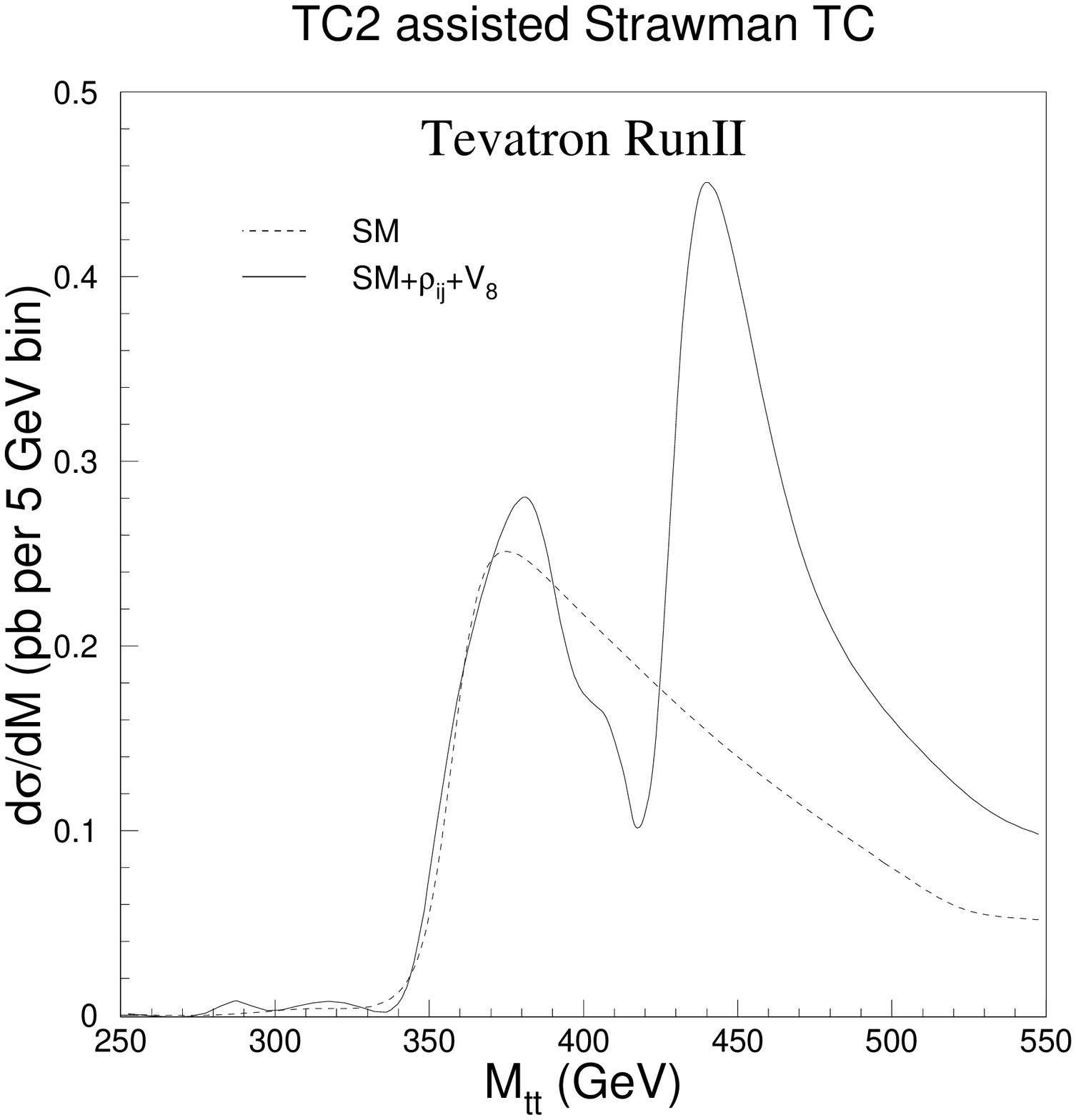}
\vskip1.8truecm
 \caption{\it
      Production of $\ol tt$ in $\ol pp$ collisions at $\ecm=2\,\tev$
      according to the TCSM model of Ref.~[27].      
    \label{fig24} }
\end{figure}

Run~II of the Tevatron Collider begins in Spring 2001. The first stage,
Run~IIa, is intended to collect $2\,\ifb$ of data with significantly enhanced
CDF and D\O\ detectors featuring new silicon tracking systems. It is planned
that, after a brief shutdown to replace damaged silicon, Run~IIb will bring
the total data sets for each detector to $15\,\ifb$ or more before the LHC is
in full swing in 2006 or so. Low--scale technicolor, if it exists, will be
discovered at one of these colliders---most likely the Tevatron!

Studies of the reach for low--scale technicolor using all the processes of
the TCSM are just beginning. Early examples do not include sophisticated
detector simulations coupled to the event
generators.\cite{tcsm_singlet,tcsm_octet,pythia} I discuss results for the
color--singlet sector first.

The expected reach of CDF in Run~IIa for the $\tro \ra W^\pm \tpi \ra
\ell^\pm \nu_\ell \ts b \ts {\rm jet}$ processes is shown in Figs.~19 and~20
for the extreme cases $M_V = M_A = 100$ and $400\,\gev$.\cite{handa}.  These
plots assume the same selections and systematic uncertainty as in the
published Run~I data,\cite{cdfwpi} but double the signal efficiency (1.38\%
vs. 0.69\%). In the case of $M_V = M_A = 100\,\gev$, the $5\sigma$ discovery
reach is the same as the 95\% excluded region in Run~I, while the Run~IIa
excluded region extends up to $M_{\tro} = 240\,\gev$ and $M_{\tpi} =
135\,\gev$. When $M_V = 400\,\gev$, the $5\sigma$ discovery region extends up
to $(M_{\tro}, M_{\tpi}) = (210, 115)\,\gev$ while the excluded region
reaches to $(260,145)\,\gev$.

The reach in $\tro,\tom \ra e^+e^-$ expected by D\O\ for $M_V = 100$ and
$200\,\gev$ and other TCSM parameters (see above) is shown in
Fig.~21.\cite{meena} Recall that the sensitivity to this process increases as
$M_V$ does. As long as $Q_U + Q_D = \CO(1)$, masses $M_{\tro,\tom}$ up to
450--500~GeV should be accessible in the $e^+e^-$ channel.

The ATLAS Collaboration has studied its reach for $\tro \ra W^\pm Z, \ts
W^\pm \tpi, \ts Z \tpi$ and for $\tom \ra \gamma\tpi$.\cite{atlastdr}
Figure.~22 shows $\tropm \ra W^\pm Z \ra \ell^\pm \nu_\ell \ell^+\ell^-$ for
several $\tro$ and $\tpi$ masses and a luminosity of $10\,\ifb$. Detailed
studies have not been published for this and other modes in which the other
TCSM parameters are varied.  Still, it is clear from Fig.~22 that the higher
energy and luminosity of the LHC ought to make it possible to completely
exclude, or discover, low--scale technicolor for any reasonable set of TCSM
parameters.

Finally, and very briefly, I turn to the prospects for studying the
color--nonsinglet sector of low--scale technicolor at the Tevatron and LHC;
see Ref.~[27] (TCSM--2). As I mentioned, the simplest implementation of TC2
models requires two color $SU(3)$ groups, one that is strongly--coupled at
1~TeV for the third generation quarks $(t,b)$ and one that is weakly--coupled
for the two light generations. These two color groups must be broken down to
the diagonal $SU(3)$ near 1~TeV, and this remaining symmetry is identified
with ordinary color. The most economical way to achieve this is to have two
doublets of technifermions $T_1= (U_1,D_1)$ and $T_2 = (U_2,D_2)$, which
transform respectively as $(3,1,\Ntc)$ and $(1,3,\Ntc)$ under the two color
groups and technicolor. They condense with each other to achieve the desired
breaking to $SU(3)_C$.\cite{tctwokl}

The main phenomenological consequence of this scenario for TC2 breaking is
that the $\suc$ gluons mix with an octet of massive ``colorons'', $V^A_8$ ($A
= 1,\dots,8$), the gauge bosons of the broken topcolor $SU(3)$, and with four
color--octet technirhos $\rho^A_{ij}\sim \ol T_i \lambda_A T_j$ ($i, j =
1,2$).\cite{tcsm_octet} The colorons decay strongly to top and bottom quarks
and weakly to the light quarks~\cite{tctwohill}. Alternatively, there is a
flavor--universal variant of TC2~\cite{ccs} in which colorons decay with
equal strength to all quark flavors. In TCSM--2, we assume for simplicity
that all $\rho_{ij}$ are too light to decay to pairs of
technipions.~\footnote{The colored technipion sector of a TC2 model is bound
to be very rich. Thus, it is not clear how the limits on leptoquarks
discussed above are to be interpreted. This is work for the future.} Then,
they decay (via gluon and coloron dominance) into $\ol q q$ and $gg$ dijets
and into $g \octpi$ and $g\pi_{T1}$.

Even this minimal, simplified TC2 version of the TCSM has a much richer set
of dijet spectra and other hadron collider signals than the one--family model
discussed above.\cite{multiklrm,cdfjets}. We are just beginning to study
it. Some preliminary examples of dijet production based on the assumptions of
TCSM--2 are shown in Figs.~23 and 24 for $\ecm = 2\,\tev$ at the Tevatron. In
both figures the coloron mass is 1.2~TeV while the input $\troct$ masses
range from 350 to 500~GeV.\footnote{The pole masses are shifted somewhat from
these input values by mixing effects.} Figure~23 shows $\ol b b$ production
with a strong resonance at 300~GeV (i,e., below $\ol t t$ threshold).
Figure~24 shows $\ol t t$ production with roughly a factor of two enhancement
over the standard model. Both signals are ruled out by Run~I measurements of
the $\ol b b$ and $\ol t t$ cross sections.

Many more studies of both the color--singlet and nonsinglet sectors of the
TCSM need to be carried out. An ongoing Run~II workshop studying strong
dynamics at Fermilab will begin this before Run~II starts next
March. Presumably, the CDF and D\O\ collaborations will carry out
detector--specific simulations in the next year or two. Meanwhile, we can
expect more detailed, and more incisive, studies from the LEP collaborations
to appear. And we hope that ATLAS and CMS will consider more thoroughly the
possibility of strong dynamics beyond the standard model before they begin
their runs later in the decade.

\section*{Acknowledgements}

I wish to express my appreciation for a very stimulating conference at La
Thuile~2000 to the organizers Giorgio Belletini, Mario Greco, Giorgio
Chiarelli and to Claudia Tofani and other members of the marvelous
secretariat. I am indebted to my colleagues, members of the Run~II Strong
Dynamics Workshop and others for their help and constructive comments
throughout the course of this work. In particular, I thank Georges Azuelos,
Sekhar Chivukula, Estia Eichten, Andre Kounine, Greg Landsberg, Richard Haas,
Takanobu Handa, Robert Harris, Kaori Maeshima, Meenakshi Narain, Steve
Mrenna, Steve Parke, Francois Richard, Elizabeth Simmons, and John
Womersley. This research was supported in part by the U.~S. Department of Energy
under Grant~No.~DE--FG02--91ER40676.

\end{document}